\documentclass[%
 reprint,
nofootinbib,
 amsmath,amssymb, amsfonts,
 showkeys,
prl
]{revtex4-2}

\usepackage{graphicx}% Include figure files
\usepackage{dcolumn}% Align table columns on decimal point
\usepackage{bm}% bold math
\usepackage[dvipsnames]{xcolor} % ADDITIONAL COLORS
\usepackage[
    colorlinks,
    citecolor = RoyalBlue,
    linkcolor = RoyalBlue,
    urlcolor = RoyalBlue,
]{hyperref}

\usepackage{layouts}
\usepackage{float}
\usepackage{placeins}

\newcommand\thefontsize{The current font size is: \f@size pt}

\begin{document}

% \preprint{APS/123-QED}

\title{Exceptional Bound States in the Continuum}% Force line breaks with \\

\author{Adri\`a Can\'os Valero,$^{1}$ Zoltan Sztranyovszky,$^{2}$ Egor A. Muljarov,$^3$ Andrey Bogdanov,$^{4,5}$ and Thomas Weiss$^{1,6}$}
\affiliation{$^{1}$Institute of Physics, University of Graz, and NAWI Graz, 8010 Graz, Austria }
\affiliation{$^2$ School of Chemical Engineering, University of Birmingham, Birmingham B15 2TT, United Kingdom}
\affiliation{$^3$ School of Physics and Astronomy, Cardiff University, Cardiff CF24 3AA, United Kingdom}
\affiliation{$^{4}$Qingdao Innovation and Development Center of Harbin Engineering University, Sansha Road 1777, Qingdao 266404, China}
\affiliation{$^5$School of Physics and Engineering, ITMO University, Kronverkskiy pr. 49, 197101, St. Petersburg, Russia}
\affiliation{$^{6}$ Physics Institute and SCoPE, University of Stuttgart, 70569 Stuttgart, Germany }

\begin{abstract}

%Bound states in the continuum (BICs) and exceptional points (EPs) are raising a lot of attention in nanophotonics. BICs are non-radiating eigenmodes with the spectrum embedded in the radiation continuum of the ambient space enabling drastic enhancement of the electromagnetic field at the nanoscale. EPs are singularities appearing in non-Hermitian systems when two or more eigenmodes and their complex coalesce leading to extreme sensitivity against small perturbations. Both BICs and EPs are unique features of non-Hermitian systems that emerge from seemingly incompatible mechanisms. Here, we determine the conditions for which an arbitrary number of BICs can coalesce into an EP. This EP-BIC inherits properties from both singularities;  it does not radiate and becomes extremely sensitive to perturbations. We validate our theory with numerical simulations of stacked dielectric metasurfaces with second and third-order EP-BICs. Breaking the symmetry of the system, we show that the dependence of the loss for EP-BIC behaves in an unusual way not corresponding to the asymptotics of the conventional quasi-BICs.

Bound states in the continuum (BICs) and exceptional points (EPs) are unique singularities of non-Hermitian systems. BICs demonstrate enhancement of the electromagnetic field at the nanoscale, while EPs exhibit high sensitivity to small perturbations. Here, we demonstrate that several BICs can be merged into one EP, forming an EP-BIC. The resulting state inherits properties from both BICs and EP, namely, it does not radiate and shows extremely high sensitivity to perturbations.  We validate the developed theory with numerical simulations and demonstrate the formation of second and third-order EP-BICs in stacked dielectric metasurfaces. We also show that the losses of the resulting leaky resonances exhibit an anomalous behavior when the unit cell symmetry is broken, which differs from the asymptotics commonly attributed to BICs.
\end{abstract}

%Use showkeys class option if keyword
\maketitle

%\tableofcontents

% \printinunitsof{cm}\prntlen{\textwidth}
% the font size is \thefontsize pt

Open systems inherently imply interaction with the surrounding space via the exchange of energy. They can be found in both classical and quantum systems, involving acoustical or optical waves~\cite{ashida2020non,el2018non,el2019dawn}. Formally, open systems can be described with non-Hermitian Hamiltonians \cite{feshbach1958unified,lindblad1976generators,lee2019topological,song2019non}. In the past few years, this general framework has revealed new physical phenomena and functionalities that are currently under intense exploration, for instance, coherent perfect absorption~\cite{chong2010coherent}, unidirectional energy transport~\cite{regensburger2012parity}, single-mode lasing~\cite{liu2017integrated}, static nonreciprocity~\cite{coulais2017static} or superscattering \cite{CanosValero2023}.

One of the most remarkable consequences of non-hermiticity is the emergence of {\it exceptional points} (EPs) in the eigenenergy spectra. EPs are spectral singularities where at least two eigenvectors and their associated eigenvalues coalesce~\cite{Miri2019Jan}. In close proximity to an EP, the eigenvalues exhibit a strongly enhanced sensitivity to perturbations, making them prospective for sensing applications~\cite{heiss2012physics,wiersig2014enhancing}, only limited by noise \cite{langbein2018no}. The coalescence of several eigenvectors results in the formation of higher-order EPs showing even higher sensitivity to perturbations~\cite{Hodaei2017Aug}. In optics, EPs have been primarily investigated in $\mathcal{PT}$-symmetric systems \cite{Miri2019Jan}. However, they can also emerge in passive structures, such as plasmonic metasurfaces~\cite{song2021plasmonic,xu2023subwavelength,park2020symmetry,sang2021tuning}, microresonators \cite{netherwood2023exceptional} and waveguides~\cite{wiersig2014enhancing,guo2009observation}, or even single nanoparticles~\cite{valero2024bianisotropic}.

Another fascinating type of singularity of open systems are {\it bound states in the continuum} (BICs)~\cite{hsu2016bound,Koshelev2023May,zhen2014topological}. BICs were first predicted in quantum mechanics \cite{von1929some,stillinger1975bound}, but today they are actively studied in photonics, acoustics, and hydrodynamics \cite{hsu2016bound,azzam2021photonic,plotnik2011experimental}. They are non-radiating resonances, despite being embedded in the continuum of radiating waves of the environment. Therefore, their radiative Q factor is infinite, while the total Q factor can be finite due to material losses~\cite{Liang2020Sep,Azzam2018Dec}. 

The radiation cancellation can arise due to symmetry or by tuning the parameters of the system such that radiation from all open channels is suppressed. As a result, BICs are classified as symmetry protected (S-BICs) or accidental, respectively \cite{hsu2016bound,neale2021accidental}. In photonics, S-BICs have been mainly studied in periodic metasurfaces. Symmetry breaking turns a S-BIC into a quasi-BIC with a finite radiative Q factor, enabling precise control of the radiative losses~\cite{koshelev2018asymmetric}, resulting in giant amplification of the incident radiation. Today, BICs present exciting opportunities for the development of compact high-Q platforms for biosensing, polaritonics, and nonlinear nanophotonics~\cite{Yesilkoy2019Jun, Kravtsov2020Apr,Maggiolini2023Aug}.

Given the unusual properties of both kinds of singularities, an intriguing question to ask is  {\it 'Can two or more BICs coalesce into an EP?'}. Such ``EP-BIC'' may possess hybrid characteristics inherited from the two singularities, that could lead to unexpected behavior and may prove beneficial for applications. Surprisingly, there is still no answer to this question, despite several reports of systems displaying BICs and EPs in the same parameter space~\cite{yang2020unconventional,deng2022extreme,sakotic2023non,qin2022exceptional,kikkawa2020bound}.

In this study, we find a general recipe for the coalescence of several BICs into an EP of arbitrary order, forming an EP-BIC. After establishing a coupling theory valid for eigenmodes with small radiation losses, we verify our results numerically by designing a bilayer and a trilayer dielectric metasurface operating in the visible range, realizing second and third order EP-BICs. Remarkably, the novel states retain the non-radiating behavior of BICs, while simultaneously exhibiting the square and cubic root dispersion of EPs. Furthermore, under small symmetry-breaking perturbations, the losses of the resulting quasi-BICs no longer follow the conventional asymptotics for the Q factor vs asymmetry parameter~\cite{koshelev2018asymmetric,aigner2022plasmonic,tao2023tunable}.

\begin{figure}[t]
\includegraphics[width=0.9\linewidth]{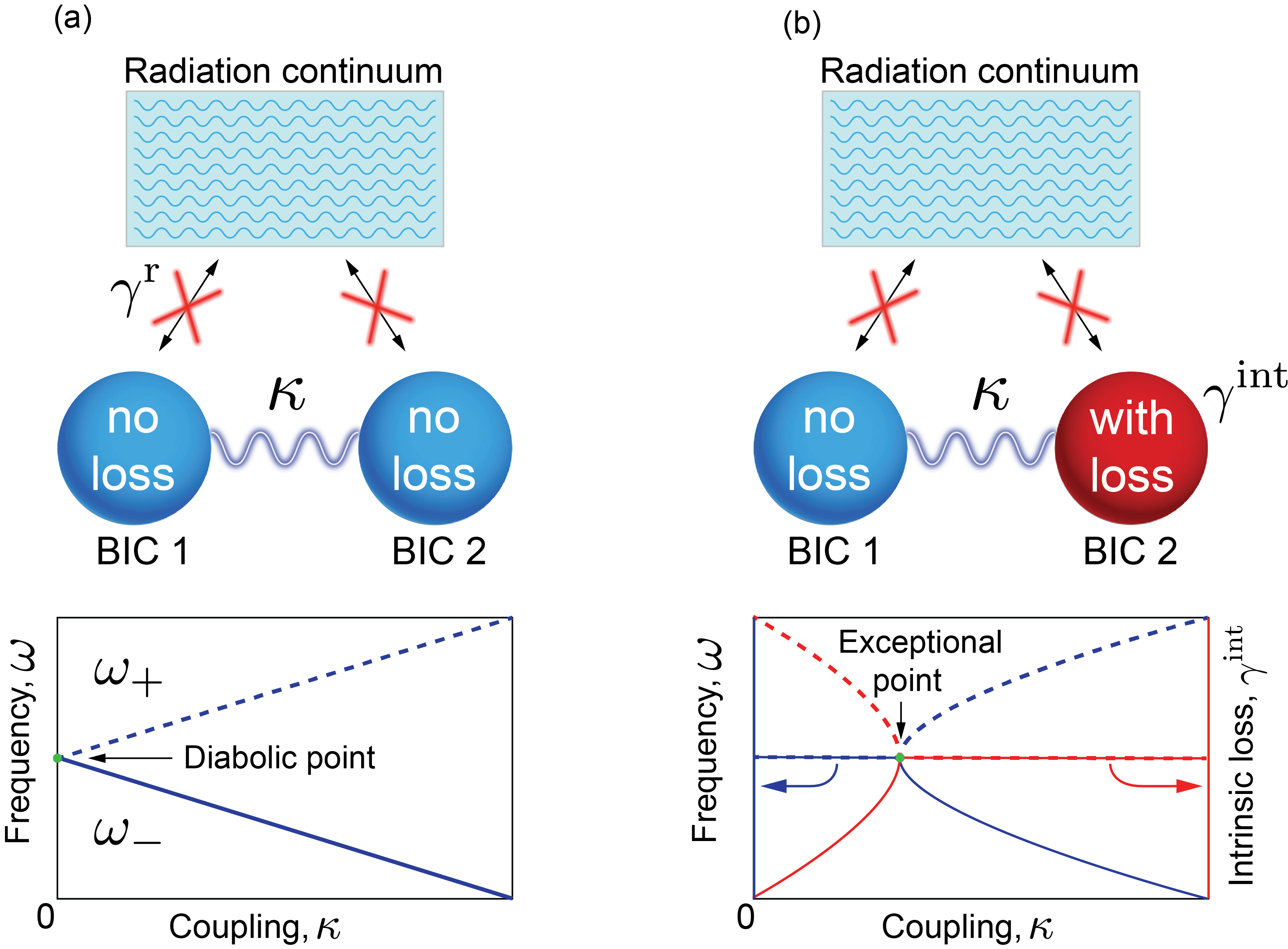}% Here is how to import EPS art
\caption{\label{fig1} Exceptional bound states in the continuum. (a) Upper panel: Illustration of two  BICs  without intrinsic losses, coupled in the near field by $\kappa$. The BICs are characterized by the complete suppression of radiation losses $\gamma^\text{r}$ (infinite radiative Q factor), and have identical eigenfrequencies $\omega_1=\omega_2$ when $\kappa=0$. Lower panel: evolution of the resonance frequencies of the BICs vs. $\kappa$. (b) Same as (a), when BIC 2 has $\gamma^{\text{int}}\neq 0$. When $\kappa=\gamma^{\text{int}}/2$, this results in the formation of an EP-BIC with infinite radiative Q factor. }
\end{figure}

We gain initial insight into the formation of an EP-BIC, by making use of an effective Hamiltonian formalism to qualitatively describe the behavior of the eigenmodes of the open system, which are also referred to as resonant states or quasinormal modes~\cite{both2021resonant,kristensen2020modeling,lalanne2018light}.  BICs and EPs can only coexist in the same parameter space if at least two modes are present. Therefore,  we write the simplest possible model effective Hamiltonian \cite{sadreev2021interference,shabanov2009resonant}, representing two coupled, radiative resonators (sketched in Fig.\ref{fig1}):
\begin{equation} \label{effective0}
    \hat{H}=
    \begin{pmatrix}
    \tilde{\omega}_1&\tilde{\kappa}\\
    \tilde{\kappa}&\tilde{\omega}_2
    \end{pmatrix}.
\end{equation}
$\hat{H}$ features complex eigenfrequencies of the form $\tilde{\omega}_{1,2}=\omega_{1,2}-i(\gamma_{1,2}^{\text{r}} + \gamma_{1,2}^{\text{int}})$, and coupling coefficients $\tilde{\kappa}=\kappa-i\sqrt{\gamma_1^{\text{r}}\gamma_2^{\text{r}}}$. $\omega_{1,2}$,  $\gamma_{1,2}^{\text{r}}$ and $\gamma_{1,2}^{\text{int}}$ are, respectively, the resonance frequencies, radiation and intrinsic losses of the uncoupled modes, while $\kappa$ is the near field coupling coefficient, here assumed to be real. If the mode is allowed to radiate, the total losses are half the linewidth of the resonant peaks. Equation~(\ref{effective0}) does not specify the physical nature of the system, which can be photonic, or acoustic. The eigenfrequencies $\tilde{\omega}_{\pm}$ are the solutions to the dispersion relation:
\begin{equation} \label{dispersion}
    \tilde{\omega}_{\pm}=\frac{\tilde{\omega}_1+\tilde{\omega}_2}{2}\pm\Lambda,
\end{equation}
where $\Lambda = \sqrt{(\tilde{\omega}_1-\tilde{\omega}_2)^2/4+\tilde{\kappa}^2}$. In the first stage, we set $\gamma_{1,2}^{\text{int}}=0$ (no intrinsic loss). EPs as well as the two known BIC types, accidental and S-BICs, can arise in this system. When $\Lambda=0$ , equation~(\ref{dispersion}) tells us that the modes coalesce at an EP at the complex frequency $\tilde{\omega}_{\text{EP}}=(\tilde{\omega}_1+\tilde{\omega}_2)/2$.  Similarly, an accidental BIC may occur when  $\kappa(\gamma^{\text{r}}_1-\gamma^{\text{r}}_2)=\sqrt{\gamma^{\text{r}}_1\gamma^{\text{r}}_2}(\omega_1-\omega_2)$ \cite{hsu2016bound}. Finally, S-BICs can be found trivially by setting $\gamma^{\text{r}}_1=\gamma^{\text{r}}_2=0$, decoupling both modes from the radiation continuum. In the third case, $\hat{H}$ is purely Hermitian and the eigenspectrum hosts no EPs. This last scenario is illustrated in Fig.~\ref{fig1}(a), which depicts two identical resonant structures, each supporting a hypothetical S-BIC with identical resonance frequencies. The lower panel in Fig.~\ref{fig1}(a) shows the resonance frequencies of the two BICs calculated as a function of $\kappa$. When $\kappa = 0$,  (uncoupled resonators) the two modes form a diabolic point \cite{doiron2022realizing,teller1937crossing}. With increasing $\kappa$, the resonance frequencies split, with a gap proportional to $\kappa$. This is the conventional behavior of S-BICs, which was recently experimentally reported in Ref.~\cite{doiron2022realizing}. 

\begin{figure*}[!htb]
\includegraphics[width=0.9\textwidth]{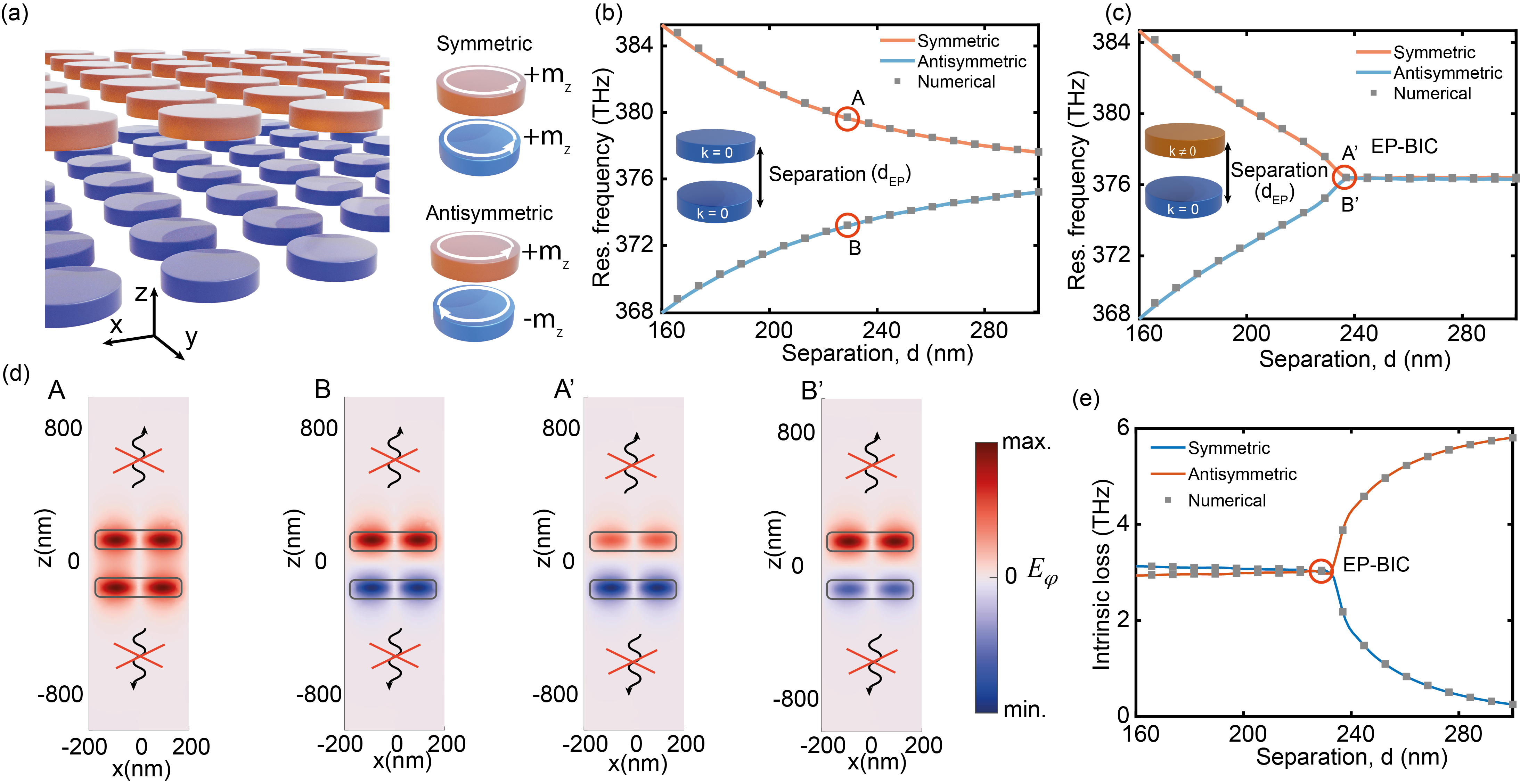}% Here is how to import EPS art
\caption{\label{fig2} Formation of an EP-BIC in a bilayer dielectric metasurface. (a) Left: artistic view of the design, consisting of two stacked metasurfaces of dielectric disks  with an identical radius of 150~nm,  height 50~nm, and period 400~nm . The refractive index of the top metasurface is $n_t=4+ik$, while it is $n_b=4$ in the bottom one. Right: symmetric  and antisymmetric BICs formed by the hybridization of the MD BICs of each metasurface.  (b) Comparison between the numerical and analytical results for the resonance frequencies vs. separation, with $k=0$ everywhere. (c) Same as (b), but with $k= 0.07$ in the top metasurface. (d)~Azimuthal component of the electric field for the two BICs at points A,B, A' and B' indicated in (b) and (c). (e) Half linewidth of the BICs vs. separation, for the case in (c).  }
\end{figure*}

Despite this, an eigenmode of Eq.~(\ref{effective0}) cannot simultaneously satisfy the BIC and EP conditions, independently of the number of tuning parameters. To understand why, consider the trace of $\hat{H}$, which must remain invariant under any basis change, and reads $\mathrm{tr}(\hat{H}) = \tilde{\omega}_1+\tilde{\omega}_2$. At an EP, both eigenmodes must have the same eigenfrequency $A$, so that $\mathrm{tr}(\hat{H}) =\tilde{\omega}_1+\tilde{\omega}_2 =2A$. Since BICs cannot radiate, $A$ must be real. The only solution is $\gamma^{\text{r}}_1=\gamma^{\text{r}}_2=0$, i.e. two S-BICs at a diabolic point. This argument is insufficient for a larger number of modes. For instance, in a $3\times 3$ Hamiltonian, two BICs might form an EP, while another mode $B$ could be leaky, such that  $\mathrm{tr}(\hat{H}) =\tilde{\omega}_1+\tilde{\omega}_2+\tilde{\omega}_3=2A+\tilde{B}$ , where $\tilde{B}$ can be complex. Hence, mode $B$ could 'keep' all the losses, i.e., $\gamma^{\text{r}}_B=\gamma^{\text{r}}_1+\gamma^{\text{r}}_2+\gamma^{\text{r}}_3$. Surprisingly, we demonstrate in the Supplemental Material \cite{Suppl_Mat_1} that, irregardless of the number of modes, BICs (S-BICs or accidental) in a purely radiative system can only form diabolic points. 

% First, the coupling between the two modes can be made assymmetric, i.e. $\kappa_{12}\neq \kappa_{21}$. When the uncoupled modes are degenerate ($\omega_1=\omega_2$), the eigenfrequencies are $\omega_{\pm}=\omega_1\pm\sqrt{\kappa_{12}\kappa_{21}}$. In the limit of unidirectional coupling, two S-BICs coalesce at an EP-BIC. Although exotic, such coupling configuration does not necessarily violate reciprocity \cite{zhong2020hierarchical,wang2019arbitrary}.

The restriction can be lifted by introducing intrinsic (dissipation) losses, $\gamma^{\text{int}}$, into one of the S-BICs, as illustrated in Fig.~\ref{fig1}(b). Now, one of the S-BICs has a nonzero imaginary part (yet still remains nonradiative).  Setting $\Lambda=0$ leads to the EP conditions $2\kappa=\pm\gamma^{\text{int}}$. Figure~\ref{fig1}(b) (lower panel) shows the resonance frequencies vs. $\kappa$ for this case, with $\kappa>0$. Note that the only difference between  Figs.~\ref{fig1}(a) and ~\ref{fig1}(b) is the addition of intrinsic loss into one of the BICs. Instead of a linear dependence in the strong coupling regime, the BICs collapse at an EP-BIC at the onset from weak to strong coupling. Note that this approach is also applicable to accidental BICs.

Next, we will demonstrate the formation of an EP-BIC in a real system. The structure we investigate consists of two stacked high-index dielectric metasurfaces in vacuum [illustrated in Fig.~\ref{fig2}(a)], with identical dimensions, listed in the caption of Fig.~\ref{fig2}. Each isolated metasurface is designed to support a S-BIC at normal incidence, at the resonance frequency $\omega_0$. This particular S-BIC can be pictured as an array of out-of-plane magnetic dipole moments. When placed close to each other, the BICs hybridize, forming a pair of symmetric and antisymmetric modes, schematically shown in the right panels of Fig.~\ref{fig2}(a). The top metasurface can be made lossy, introducing a small imaginary part $\gamma^{\text{int}}$ to the corresponding BIC. Hence, the BIC retains its nonradiative nature, but has a finite Q-factor $Q=\omega_0/2\gamma^{\text{int}}$.  To obtain the Hamiltonian $\hat{H}_M$, we derived a coupling mode theory using the BICs of the individual metasurfaces as a basis~\cite{Suppl_Mat_1,tao2020coupling}. 
When $\gamma^{\text{int}}\ll\omega_0$ and assuming small interaction , $ \hat{H}_{\text{M}}$ simplifies to \cite{Suppl_Mat_1}:
\begin{equation} \label{hamiltonSimple}
    \hat{H}_{\text{M}} \approx \begin{pmatrix}
        \omega_0-i\gamma^{\text{int}} & -g\omega_0\\
        -g\omega_0 & \omega_0
    \end{pmatrix}.
\end{equation}
Here, $g$ can be retrieved analytically from the fields of the uncoupled BICs as $g = \int_{V_{\text{t}}}\tilde{\mathbf{E}}_{\text{t}}(\mathbf{r})\cdot\Delta\varepsilon\cdot\tilde{\mathbf{E}}_{\text{b}}(\mathbf{r})dV,$ where the subscripts (t,b) denote the BICs from the top and bottom metasurface, respectively, and $\Delta\varepsilon=\varepsilon_0(\varepsilon_{\text{t}}-1)$ is the difference between the permittivity of the disks in the top metasurface and vacuum. The integral runs over the volume $V_{\text{t}}$ of the top disk.  Equation~(\ref{hamiltonSimple}) allows predicting the eigenfields and eigenfrequencies of the coupled BIC metasurfaces with the sole knowledge of the eigenfields in the bare constituents. This is confirmed by the good agreement between the numerical and analytical results in  Fig.~\ref{fig2}. Importantly, $\hat{H}_\text{M}$ takes the form of the toy model in Eq.~(\ref{effective0}), with $-g\omega_0$ playing the role of $\tilde{\kappa}$.  

Figure~\ref{fig2}(b) shows the results when ohmic losses are absent. With decreasing separation $d$ between the metasurfaces, the overlap between the evanescent tails of the BICs is larger, and $g$ increases. The resonance frequencies follow the same qualitative picture from the toy model in Fig.~\ref{fig1}(a). Figure~\ref{fig2}(c) and \ref{fig2}(e)  display, respectively, the resonance frequencies and dissipation losses of the BICs when the refractive index of the top metasurface is made complex. The EP-BIC emerges when $d_{\text{EP}}\approx 236$~nm, for the chosen geometrical parameters. Near the singularity, the resonance frequencies show a drastic difference with respect to the conventional scenario. For $d<d_{\text{EP}}$, they exhibit the characteristic square-root dispersion of an EP . The same can be observed with the intrinsic losses for $d>d_{\text{EP}}$. The resonance frequencies of the symmetric and antisymmetric BICs are quasi-degenerate for  $d>d_{\text{EP}}$, a phenomenon known as a bulk Fermi arc~\cite{zhou2018observation}.  Despite the presence of dissipation losses, the eigenmodes are still BICs, since their field profiles are incompatible with the electromagnetic continuum. This can be visually confirmed from the absence of propagating waves in the field distributions of the BICs [Fig.~\ref{fig2}(d)], which remain confined in the metasurfaces. At the same time, owing to the collapse of the eigenspectra, near the EP-BIC the fields of the two modes become almost identical [panels A', B' in Fig.~\ref{fig2}(d)].
\begin{figure}[!htb]
\includegraphics[width=0.45\textwidth]{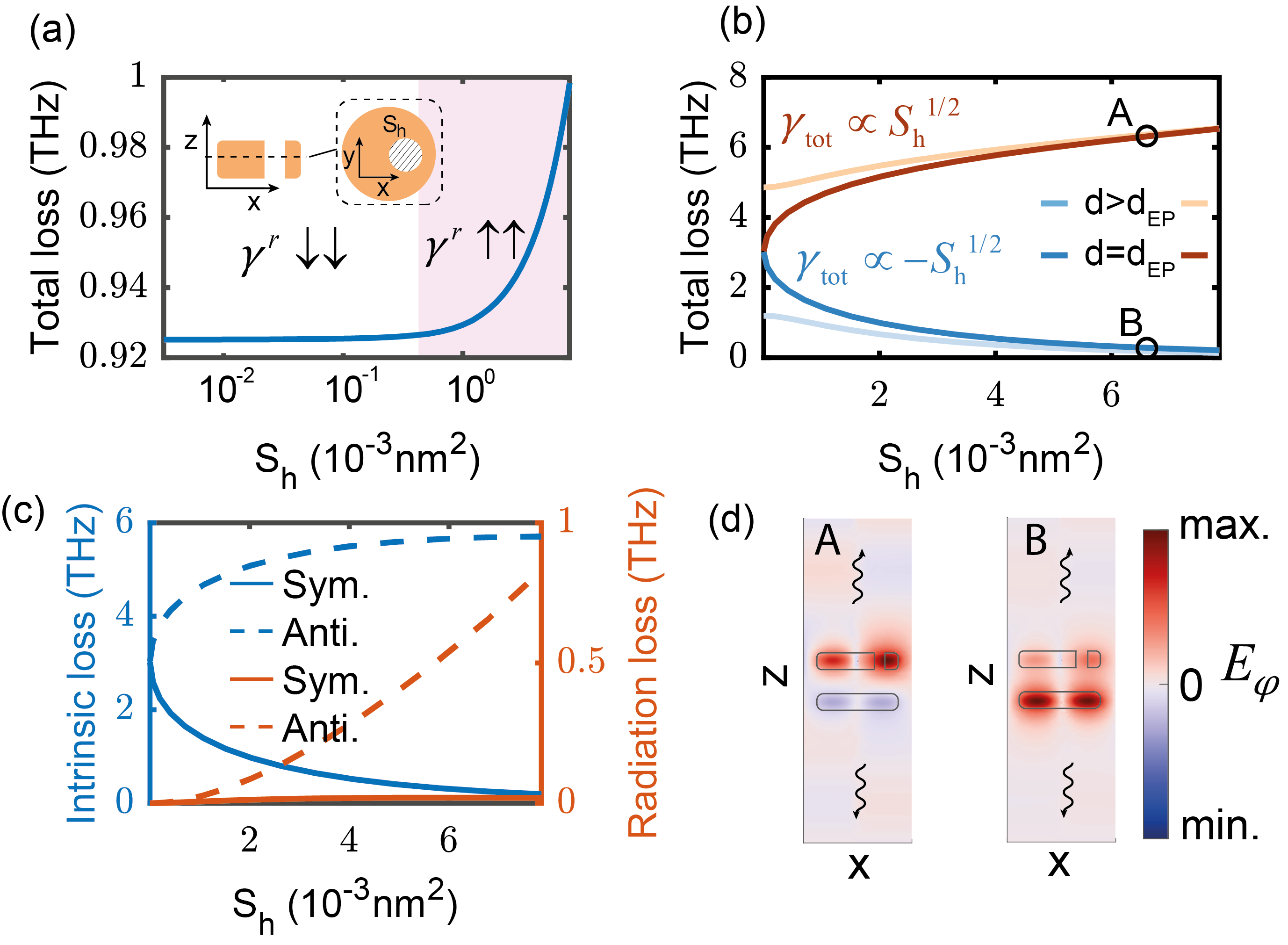} \setlength{\belowcaptionskip}{-8pt} 
\caption{\label{fig3} (a) Evolution of the total losses of a single (lossy) metasurface with same parameters as Fig.\ref{fig2}, when etching a hole of surface $S_{\text{h}}$, offset 40~nm from the center of the disk, (see inset). The x-axis is shown in log scale. (b) Bilayer metasurface with $d=250\:\text{nm}>d_{\text{EP}}$ and  $d=236\:\text{nm}\approx{}d_{\text{EP}}$.  Orange and blue lines correspond to the antisymmetric and symmetric BICs, respectively. (c) Analysis of the radiative and intrinsic contributions to the total loss when perturbing the EP-BIC. For small perturbations, the intrinsic loss dominates and is responsible for the enhanced response. (d) Azimuthal component of the electric field for the two quasi-BICs at points A,B indicated in (b).  }
\end{figure}

Since they are nonradiative, EP-BICs cannot be optically probed from the far field. We now study the effect of symmetry-breaking perturbations that introduce radiation losses into the system. As a result, the EP-BIC condition is violated, and the perturbed modes turn into radiative quasi-BICs. We demonstrate that the linewidths of the resonant peaks no longer follow the behavior attributed to S-BICs up to date. 

As sketched in the inset of Fig.~\ref{fig3}(a), we introduced a small off-center hole with cross sectional area $S_\text{h}$ in the top metasurface, which breaks the in-plane symmetry and couples the S-BICs to normally incident plane waves. We first analyzed the behavior of the lossy S-BIC in a single metasurface [Fig.~\ref{fig3}(a)]. With increasingly large $S_{\text{h}}$, two regimes can be distinguished. Small holes lead to a linear growth of $\gamma$, in agreement with first order perturbation theory \cite{weiss2016dark}, where absorption losses dominate \cite{aigner2022plasmonic,tao2023tunable}. With larger $S_{\text{h}}$, in accordance with second order perturbation theory~\cite{koshelev2018asymmetric}, the BICs interact with radiative modes, resulting in radiative losses. 
% This behavior is in agreement with all previous predictions and observations of quasi-BICs in passive structures~\cite{koshelev2018asymmetric,azzam2018formation,overvig2020selection,gorkunov2020metasurfaces,kang2022coherent,zhou2022dual,blanchard2016fano,deriy2022bound}. 
This is in striking contrast with the behavior near the EP-BIC shown in Fig.~\ref{fig3}(b). Here, instead, a square-root dependence with small $S_{\text{h}}$ is observed. Far from the EP-BIC, the quasi-BICs recover the linear dependence. This suggests a non-trivial behavior of the intrinsic loss. Interestingly, while the antisymmetric mode becomes more lossy, the symmetric mode linearly decreases its loss. These unusual characteristics can be observed in the reflection spectra for $y$ polarized light~\cite{Suppl_Mat_1}. We also confirmed a similar qualitative behavior in reciprocal space, when the symmetry is broken by changing the Bloch wavevector~\cite{Suppl_Mat_1}.

Next, we show that the observed trends can be explained by taking into account the combined effects of a second order EP and the BICs in the perturbation series. Consider the discriminant $\Lambda$ in Eq~(\ref{dispersion}). At the EP-BIC, $\Lambda=0$. Assume now that we modify the losses $\delta\gamma$ of one of the BICs. This represents the effect of symmetry breaking. To leading order, $\Lambda\approx i\sqrt{\gamma^{\text{int}}\delta\gamma/2}$~\cite{Suppl_Mat_1}, confirming the square-root assymptotics near the EP. However, to first order the losses of the BIC on the top metasurface follow the law $\delta\gamma=2aS_{\text{h}}$, where $a$ is a constant. Inserting this Ansatz into the expansion of $\Lambda$, and plugging the latter in Eq.~(\ref{dispersion}), we arrive at the dispersion relation near the EP-BIC:
\begin{equation} \label{EP-BIC dispersion}
    \tilde{\omega}_{\pm}=\tilde{\omega}_{\text{EP}}-iaS_{\text{h}}\pm i\left(\gamma^{\text{int}}\right)^{1/2}a^{1/2}S_{h}^{1/2}.
\end{equation}
Equation~(\ref{EP-BIC dispersion}) features a linear and a square-root contribution to $\gamma^{\text{tot}}$. The square-root term is absent in conventional BICs, and appears due to the EP dispersion. The different signs explain the increase and decrease in the losses of the two quasi-BICs for small $S_{\text{h}}$.  However, the analysis does not reveal if the new features are connected to absorption and/or radiation. These two contributions can always be separated in weakly dispersive media~\cite{lalanne2018light,Suppl_Mat_1}. 
The intrinsic losses can be calculated by taking the ratio between half the absorbed power and the electromagnetic energy in the unit cell volume~\cite{Suppl_Mat_1}. The radiation losses can be found as $\gamma^{\text{r}}=\gamma^{\text{tot}}-\gamma^{\text{int}}$. Figure~\ref{fig3}(c)  displays the evolution of $\gamma^{\text{int}}$ and $\gamma^{\text{r}}$ for the two quasi-BICs at $d\approx d_{\text{EP}}$. It can be confirmed that only $\gamma^{\text{int}}$ is responsible for the square-root trend, as well as the decrease in the losses of the symmetric mode. Conversely, the radiation losses grow quadratically in both quasi-BICs, but saturate very fast in the symmetric mode. This is because increasing $S_{\text{h}}$ localizes the symmetric mode on the bottom metasurface, leading to a smaller intrinsic loss and a weaker $S_{\text{h}}$ dependence, as can be confirmed in the field distributions of the quasi-BICs [Fig.~\ref{fig3}(d)].

\begin{figure}[htb!]
\includegraphics[width=0.4\textwidth]{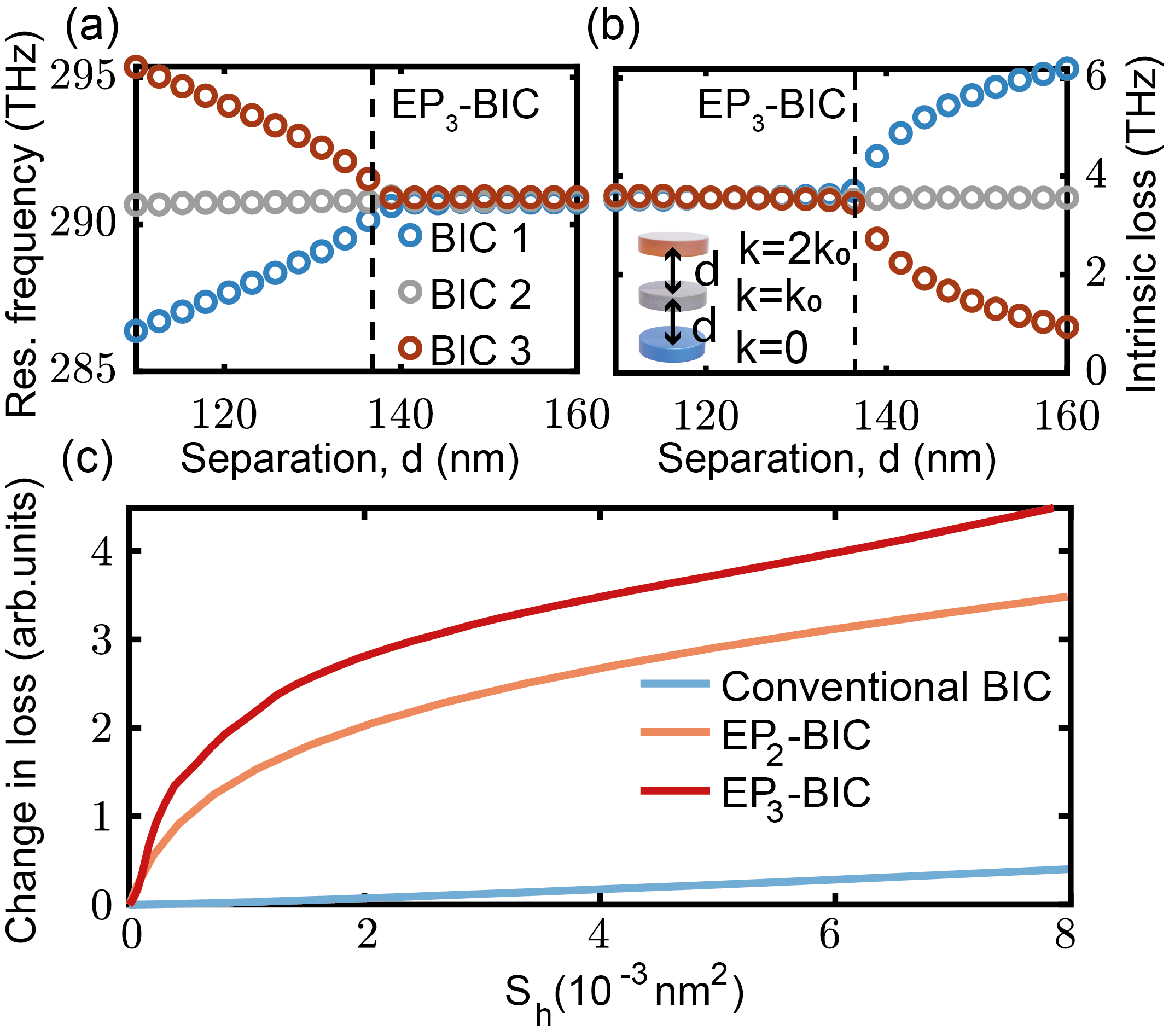} 
\setlength{\belowcaptionskip}{-8pt}\caption{\label{fig4} Realization of third order EP-BICs ($\text{EP}_3\text{-BIC}$). (a) Evolution of the resonance frequencies and (b) the intrinsic loss of three coupled S-BICs in a trilayer metasurface, (unit cell in the inset). The refractive indices from top to bottom are $n=4+ik$,  ($k_0=0.05$) . (c) Linewidth changes for conventional S-BICs and EP-BICs with $m=2,3$. Holes are etched on the top (lossy) metasurface for all cases. Geometrical parameters for all calculations: Disks of height 150~nm, radius 150~nm. Period: 350~nm.}
\end{figure}

 The methodology outlined here can be extended to EP-BICs of arbitrary order, provided that intrinsic losses are properly distributed among the resonators. For instance, a third order EP-BIC can be realized by stacking three BIC metasurfaces, [see inset of Fig.~\ref{fig4}(b)], obeying the Hamiltonian:
\begin{equation} \label{third_order}
    \hat{H}_{\text{M}} \approx \begin{pmatrix}
       \omega_0-2i\gamma^{\text{int}}& -g\omega_0&0\\
       -g\omega_0& \omega_0-i\gamma^{\text{int}}& -g\omega_0\\
       0&-g\omega_0&\omega_0
    \end{pmatrix}.
\end{equation}
Equation~(\ref{third_order}) leads to a third order EP when $\gamma^{\text{int}} = \sqrt{2}g\omega_0$. The resonance frequencies and intrinsic losses of this system are shown in Fig.~\ref{fig4}(a)-(b), demonstrating the coalescence of the three BICs into a third-order EP-BIC. 

Finally, we investigate the effect of symmetry breaking perturbations for EP-BICs of arbitrary order. Employing a generalized Newton-Puisseux series \cite{heiss2012physics,Hodaei2017Aug}, we derived a general rule for the total losses of the quasi-BICs emerging from an EP-BIC of order $m$ \cite{Suppl_Mat_1}, which reads:
\begin{equation} \label{arbitrary}
    \gamma_h =  \gamma_{\text{EP}} +\text{Im}\{\alpha^{1/m}\zeta^{h-1}\}p^{1/m},
\end{equation}
where $h = 1 ... m$ is an integer denoting each quasi-BIC spanned by the EP-BIC, $\alpha$ is a complex number, $\zeta = e^{2\pi i /m}$ , $p$ is the assymmetry parameter (e.g., $S_{\text{h}}$), and $\gamma_{\text{EP}}$ is the intrinsic loss at the EP-BIC. A third order EP-BIC results on quasi-BICs scaling as $\gamma_h\propto p^{1/3}$, implying larger changes for small $p$ with respect to conventional BICs and second order EP-BICs. This is shown in Fig.~\ref{fig4}(c), where we compare the change in loss $\gamma_h-\gamma_{\text{EP}}$ for perturbations of a conventional S-BIC and second and third order EP-BICs. 

% In order to ensure a fair comparison, the sizes and period of the metasurfaces are kept the same, and the extinction coefficient is adjusted to k_0=0.05, so that all systems start with a BIC with approximately the same Q factor

In conclusion, in this Letter we confirmed the possibility to coalesce BICs into EPs of arbitrary order, forming an EP-BIC. The key for their realization is the introduction of asymmetric loss into a system supporting two or more BICs. We have confirmed our predictions by simulations of bilayer and trilayer metasurfaces hosting S-BICs coalescing at second- and third-order EP-BICs. This mechanism might not be unique. For instance, EP-BICs might exist in the presence of unidirectional coupling~\cite{wang2019arbitrary,wang2020waveguide,wiersig2022revisiting}. This constitutes a promising research direction, since it would remove the need of intrinsic losses that limit the achievable Q factor. \\
The novel states inherit the infinite radiative Q factors of BICs, but possess the enhanced eigenvalue sensitivity of EPs. Interestingly, introducing radiation losses to an EP-BIC of order $m$, mediated by a parameter $p$, results in a change of the total losses in the order of $p^{1/m}$, unlike the linear response of conventional S-BICs. Owing to the broad applicability of our formalism, EP-BICs are expected to be a general wave phenomenon, beyond the studied optical setup. For instance, we anticipate them to play an important role in the emerging hybrid dielectric-plasmonic metasurfaces \cite{azzam2018formation,yang2017low}. Besides their fundamental interest, EP-BICs can become promising candidates for next generation sensing, imaging, and communication platforms.

% \begin{figure}[!htb]
% \includegraphics[width=0.15\textwidth]{Figures/fields_losses.eps}% Here is how to import EPS art
% \caption{\label{fig4} Predicted linewidth changes for conventional BICs and EP-BICs of second ($EP_2-BIC$) and third order ($EP_3-BIC$),  under small perturbations $p$. }
% \end{figure}

%The authors acknowledge support of 
\FloatBarrier
% \printbibliography
% \clearpage
\bibliographystyle{apsrev4-2}
\bibliography{bibliography}% Produces the bibliography via BibTeX.

\begin{thebibliography}{61}%
\makeatletter
\providecommand \@ifxundefined [1]{%
 \@ifx{#1\undefined}
}%
\providecommand \@ifnum [1]{%
 \ifnum #1\expandafter \@firstoftwo
 \else \expandafter \@secondoftwo
 \fi
}%
\providecommand \@ifx [1]{%
 \ifx #1\expandafter \@firstoftwo
 \else \expandafter \@secondoftwo
 \fi
}%
\providecommand \natexlab [1]{#1}%
\providecommand \enquote  [1]{``#1''}%
\providecommand \bibnamefont  [1]{#1}%
\providecommand \bibfnamefont [1]{#1}%
\providecommand \citenamefont [1]{#1}%
\providecommand \href@noop [0]{\@secondoftwo}%
\providecommand \href [0]{\begingroup \@sanitize@url \@href}%
\providecommand \@href[1]{\@@startlink{#1}\@@href}%
\providecommand \@@href[1]{\endgroup#1\@@endlink}%
\providecommand \@sanitize@url [0]{\catcode `\\12\catcode `\$12\catcode
  `\&12\catcode `\#12\catcode `\^12\catcode `\_12\catcode `\%12\relax}%
\providecommand \@@startlink[1]{}%
\providecommand \@@endlink[0]{}%
\providecommand \url  [0]{\begingroup\@sanitize@url \@url }%
\providecommand \@url [1]{\endgroup\@href {#1}{\urlprefix }}%
\providecommand \urlprefix  [0]{URL }%
\providecommand \Eprint [0]{\href }%
\providecommand \doibase [0]{https://doi.org/}%
\providecommand \selectlanguage [0]{\@gobble}%
\providecommand \bibinfo  [0]{\@secondoftwo}%
\providecommand \bibfield  [0]{\@secondoftwo}%
\providecommand \translation [1]{[#1]}%
\providecommand \BibitemOpen [0]{}%
\providecommand \bibitemStop [0]{}%
\providecommand \bibitemNoStop [0]{.\EOS\space}%
\providecommand \EOS [0]{\spacefactor3000\relax}%
\providecommand \BibitemShut  [1]{\csname bibitem#1\endcsname}%
\let\auto@bib@innerbib\@empty
%</preamble>
\bibitem [{\citenamefont {Ashida}\ \emph {et~al.}(2020)\citenamefont {Ashida},
  \citenamefont {Gong},\ and\ \citenamefont {Ueda}}]{ashida2020non}%
  \BibitemOpen
  \bibfield  {author} {\bibinfo {author} {\bibfnamefont {Y.}~\bibnamefont
  {Ashida}}, \bibinfo {author} {\bibfnamefont {Z.}~\bibnamefont {Gong}},\ and\
  \bibinfo {author} {\bibfnamefont {M.}~\bibnamefont {Ueda}},\ }\href@noop {}
  {\bibfield  {journal} {\bibinfo  {journal} {Advances in Physics}\ }\textbf
  {\bibinfo {volume} {69}},\ \bibinfo {pages} {249} (\bibinfo {year}
  {2020})}\BibitemShut {NoStop}%
\bibitem [{\citenamefont {El-Ganainy}\ \emph {et~al.}(2018)\citenamefont
  {El-Ganainy}, \citenamefont {Makris}, \citenamefont {Khajavikhan},
  \citenamefont {Musslimani}, \citenamefont {Rotter},\ and\ \citenamefont
  {Christodoulides}}]{el2018non}%
  \BibitemOpen
  \bibfield  {author} {\bibinfo {author} {\bibfnamefont {R.}~\bibnamefont
  {El-Ganainy}}, \bibinfo {author} {\bibfnamefont {K.~G.}\ \bibnamefont
  {Makris}}, \bibinfo {author} {\bibfnamefont {M.}~\bibnamefont {Khajavikhan}},
  \bibinfo {author} {\bibfnamefont {Z.~H.}\ \bibnamefont {Musslimani}},
  \bibinfo {author} {\bibfnamefont {S.}~\bibnamefont {Rotter}},\ and\ \bibinfo
  {author} {\bibfnamefont {D.~N.}\ \bibnamefont {Christodoulides}},\
  }\href@noop {} {\bibfield  {journal} {\bibinfo  {journal} {Nature Physics}\
  }\textbf {\bibinfo {volume} {14}},\ \bibinfo {pages} {11} (\bibinfo {year}
  {2018})}\BibitemShut {NoStop}%
\bibitem [{\citenamefont {El-Ganainy}\ \emph {et~al.}(2019)\citenamefont
  {El-Ganainy}, \citenamefont {Khajavikhan}, \citenamefont {Christodoulides},\
  and\ \citenamefont {Ozdemir}}]{el2019dawn}%
  \BibitemOpen
  \bibfield  {author} {\bibinfo {author} {\bibfnamefont {R.}~\bibnamefont
  {El-Ganainy}}, \bibinfo {author} {\bibfnamefont {M.}~\bibnamefont
  {Khajavikhan}}, \bibinfo {author} {\bibfnamefont {D.~N.}\ \bibnamefont
  {Christodoulides}},\ and\ \bibinfo {author} {\bibfnamefont {S.~K.}\
  \bibnamefont {Ozdemir}},\ }\href@noop {} {\bibfield  {journal} {\bibinfo
  {journal} {Communications Physics}\ }\textbf {\bibinfo {volume} {2}},\
  \bibinfo {pages} {37} (\bibinfo {year} {2019})}\BibitemShut {NoStop}%
\bibitem [{\citenamefont {Feshbach}(1958)}]{feshbach1958unified}%
  \BibitemOpen
  \bibfield  {author} {\bibinfo {author} {\bibfnamefont {H.}~\bibnamefont
  {Feshbach}},\ }\href@noop {} {\bibfield  {journal} {\bibinfo  {journal}
  {Annals of Physics}\ }\textbf {\bibinfo {volume} {5}},\ \bibinfo {pages}
  {357} (\bibinfo {year} {1958})}\BibitemShut {NoStop}%
\bibitem [{\citenamefont {Lindblad}(1976)}]{lindblad1976generators}%
  \BibitemOpen
  \bibfield  {author} {\bibinfo {author} {\bibfnamefont {G.}~\bibnamefont
  {Lindblad}},\ }\href@noop {} {\bibfield  {journal} {\bibinfo  {journal}
  {Communications in Mathematical Physics}\ }\textbf {\bibinfo {volume} {48}},\
  \bibinfo {pages} {119} (\bibinfo {year} {1976})}\BibitemShut {NoStop}%
\bibitem [{\citenamefont {Lee}\ \emph {et~al.}(2019)\citenamefont {Lee},
  \citenamefont {Ahn}, \citenamefont {Zhou},\ and\ \citenamefont
  {Vishwanath}}]{lee2019topological}%
  \BibitemOpen
  \bibfield  {author} {\bibinfo {author} {\bibfnamefont {J.~Y.}\ \bibnamefont
  {Lee}}, \bibinfo {author} {\bibfnamefont {J.}~\bibnamefont {Ahn}}, \bibinfo
  {author} {\bibfnamefont {H.}~\bibnamefont {Zhou}},\ and\ \bibinfo {author}
  {\bibfnamefont {A.}~\bibnamefont {Vishwanath}},\ }\href@noop {} {\bibfield
  {journal} {\bibinfo  {journal} {Physical review letters}\ }\textbf {\bibinfo
  {volume} {123}},\ \bibinfo {pages} {206404} (\bibinfo {year}
  {2019})}\BibitemShut {NoStop}%
\bibitem [{\citenamefont {Song}\ \emph {et~al.}(2019)\citenamefont {Song},
  \citenamefont {Yao},\ and\ \citenamefont {Wang}}]{song2019non}%
  \BibitemOpen
  \bibfield  {author} {\bibinfo {author} {\bibfnamefont {F.}~\bibnamefont
  {Song}}, \bibinfo {author} {\bibfnamefont {S.}~\bibnamefont {Yao}},\ and\
  \bibinfo {author} {\bibfnamefont {Z.}~\bibnamefont {Wang}},\ }\href@noop {}
  {\bibfield  {journal} {\bibinfo  {journal} {Physical review letters}\
  }\textbf {\bibinfo {volume} {123}},\ \bibinfo {pages} {170401} (\bibinfo
  {year} {2019})}\BibitemShut {NoStop}%
\bibitem [{\citenamefont {Chong}\ \emph {et~al.}(2010)\citenamefont {Chong},
  \citenamefont {Ge}, \citenamefont {Cao},\ and\ \citenamefont
  {Stone}}]{chong2010coherent}%
  \BibitemOpen
  \bibfield  {author} {\bibinfo {author} {\bibfnamefont {Y.}~\bibnamefont
  {Chong}}, \bibinfo {author} {\bibfnamefont {L.}~\bibnamefont {Ge}}, \bibinfo
  {author} {\bibfnamefont {H.}~\bibnamefont {Cao}},\ and\ \bibinfo {author}
  {\bibfnamefont {A.~D.}\ \bibnamefont {Stone}},\ }\href@noop {} {\bibfield
  {journal} {\bibinfo  {journal} {Physical review letters}\ }\textbf {\bibinfo
  {volume} {105}},\ \bibinfo {pages} {053901} (\bibinfo {year}
  {2010})}\BibitemShut {NoStop}%
\bibitem [{\citenamefont {Regensburger}\ \emph {et~al.}(2012)\citenamefont
  {Regensburger}, \citenamefont {Bersch}, \citenamefont {Miri}, \citenamefont
  {Onishchukov}, \citenamefont {Christodoulides},\ and\ \citenamefont
  {Peschel}}]{regensburger2012parity}%
  \BibitemOpen
  \bibfield  {author} {\bibinfo {author} {\bibfnamefont {A.}~\bibnamefont
  {Regensburger}}, \bibinfo {author} {\bibfnamefont {C.}~\bibnamefont
  {Bersch}}, \bibinfo {author} {\bibfnamefont {M.-A.}\ \bibnamefont {Miri}},
  \bibinfo {author} {\bibfnamefont {G.}~\bibnamefont {Onishchukov}}, \bibinfo
  {author} {\bibfnamefont {D.~N.}\ \bibnamefont {Christodoulides}},\ and\
  \bibinfo {author} {\bibfnamefont {U.}~\bibnamefont {Peschel}},\ }\href@noop
  {} {\bibfield  {journal} {\bibinfo  {journal} {Nature}\ }\textbf {\bibinfo
  {volume} {488}},\ \bibinfo {pages} {167} (\bibinfo {year}
  {2012})}\BibitemShut {NoStop}%
\bibitem [{\citenamefont {Liu}\ \emph {et~al.}(2017)\citenamefont {Liu},
  \citenamefont {Li}, \citenamefont {Guzzon}, \citenamefont {Norberg},
  \citenamefont {Parker}, \citenamefont {Lu}, \citenamefont {Coldren},\ and\
  \citenamefont {Yao}}]{liu2017integrated}%
  \BibitemOpen
  \bibfield  {author} {\bibinfo {author} {\bibfnamefont {W.}~\bibnamefont
  {Liu}}, \bibinfo {author} {\bibfnamefont {M.}~\bibnamefont {Li}}, \bibinfo
  {author} {\bibfnamefont {R.~S.}\ \bibnamefont {Guzzon}}, \bibinfo {author}
  {\bibfnamefont {E.~J.}\ \bibnamefont {Norberg}}, \bibinfo {author}
  {\bibfnamefont {J.~S.}\ \bibnamefont {Parker}}, \bibinfo {author}
  {\bibfnamefont {M.}~\bibnamefont {Lu}}, \bibinfo {author} {\bibfnamefont
  {L.~A.}\ \bibnamefont {Coldren}},\ and\ \bibinfo {author} {\bibfnamefont
  {J.}~\bibnamefont {Yao}},\ }\href@noop {} {\bibfield  {journal} {\bibinfo
  {journal} {Nature communications}\ }\textbf {\bibinfo {volume} {8}},\
  \bibinfo {pages} {15389} (\bibinfo {year} {2017})}\BibitemShut {NoStop}%
\bibitem [{\citenamefont {Coulais}\ \emph {et~al.}(2017)\citenamefont
  {Coulais}, \citenamefont {Sounas},\ and\ \citenamefont
  {Alu}}]{coulais2017static}%
  \BibitemOpen
  \bibfield  {author} {\bibinfo {author} {\bibfnamefont {C.}~\bibnamefont
  {Coulais}}, \bibinfo {author} {\bibfnamefont {D.}~\bibnamefont {Sounas}},\
  and\ \bibinfo {author} {\bibfnamefont {A.}~\bibnamefont {Alu}},\ }\href@noop
  {} {\bibfield  {journal} {\bibinfo  {journal} {Nature}\ }\textbf {\bibinfo
  {volume} {542}},\ \bibinfo {pages} {461} (\bibinfo {year}
  {2017})}\BibitemShut {NoStop}%
\bibitem [{\citenamefont {Can{\'o}s~Valero}\ \emph {et~al.}(2023)\citenamefont
  {Can{\'o}s~Valero}, \citenamefont {Shamkhi}, \citenamefont {Kupriianov},
  \citenamefont {Weiss}, \citenamefont {Pavlov}, \citenamefont {Redka},
  \citenamefont {Bobrovs}, \citenamefont {Kivshar},\ and\ \citenamefont
  {Shalin}}]{CanosValero2023}%
  \BibitemOpen
  \bibfield  {author} {\bibinfo {author} {\bibfnamefont {A.}~\bibnamefont
  {Can{\'o}s~Valero}}, \bibinfo {author} {\bibfnamefont {H.~K.}\ \bibnamefont
  {Shamkhi}}, \bibinfo {author} {\bibfnamefont {A.~S.}\ \bibnamefont
  {Kupriianov}}, \bibinfo {author} {\bibfnamefont {T.}~\bibnamefont {Weiss}},
  \bibinfo {author} {\bibfnamefont {A.~A.}\ \bibnamefont {Pavlov}}, \bibinfo
  {author} {\bibfnamefont {D.}~\bibnamefont {Redka}}, \bibinfo {author}
  {\bibfnamefont {V.}~\bibnamefont {Bobrovs}}, \bibinfo {author} {\bibfnamefont
  {Y.}~\bibnamefont {Kivshar}},\ and\ \bibinfo {author} {\bibfnamefont {A.~S.}\
  \bibnamefont {Shalin}},\ }\href {https://doi.org/10.1038/s41467-023-40382-y}
  {\bibfield  {journal} {\bibinfo  {journal} {Nature Communications}\ }\textbf
  {\bibinfo {volume} {14}},\ \bibinfo {pages} {4689} (\bibinfo {year}
  {2023})}\BibitemShut {NoStop}%
\bibitem [{\citenamefont {Miri}\ and\ \citenamefont
  {Al{\ifmmode\grave{u}\else\`{u}\fi}}(2019)}]{Miri2019Jan}%
  \BibitemOpen
  \bibfield  {author} {\bibinfo {author} {\bibfnamefont {M.-A.}\ \bibnamefont
  {Miri}}\ and\ \bibinfo {author} {\bibfnamefont {A.}~\bibnamefont
  {Al{\ifmmode\grave{u}\else\`{u}\fi}}},\ }\bibfield  {journal} {\bibinfo
  {journal} {Science}\ }\textbf {\bibinfo {volume} {363}},\ \href
  {https://doi.org/10.1126/science.aar7709} {10.1126/science.aar7709} (\bibinfo
  {year} {2019})\BibitemShut {NoStop}%
\bibitem [{\citenamefont {Heiss}(2012)}]{heiss2012physics}%
  \BibitemOpen
  \bibfield  {author} {\bibinfo {author} {\bibfnamefont {W.}~\bibnamefont
  {Heiss}},\ }\href@noop {} {\bibfield  {journal} {\bibinfo  {journal} {Journal
  of Physics A: Mathematical and Theoretical}\ }\textbf {\bibinfo {volume}
  {45}},\ \bibinfo {pages} {444016} (\bibinfo {year} {2012})}\BibitemShut
  {NoStop}%
\bibitem [{\citenamefont {Wiersig}(2014)}]{wiersig2014enhancing}%
  \BibitemOpen
  \bibfield  {author} {\bibinfo {author} {\bibfnamefont {J.}~\bibnamefont
  {Wiersig}},\ }\href@noop {} {\bibfield  {journal} {\bibinfo  {journal}
  {Physical review letters}\ }\textbf {\bibinfo {volume} {112}},\ \bibinfo
  {pages} {203901} (\bibinfo {year} {2014})}\BibitemShut {NoStop}%
\bibitem [{\citenamefont {Langbein}(2018)}]{langbein2018no}%
  \BibitemOpen
  \bibfield  {author} {\bibinfo {author} {\bibfnamefont {W.}~\bibnamefont
  {Langbein}},\ }\href@noop {} {\bibfield  {journal} {\bibinfo  {journal}
  {Physical Review A}\ }\textbf {\bibinfo {volume} {98}},\ \bibinfo {pages}
  {023805} (\bibinfo {year} {2018})}\BibitemShut {NoStop}%
\bibitem [{\citenamefont {Hodaei}\ \emph {et~al.}(2017)\citenamefont {Hodaei},
  \citenamefont {Hassan}, \citenamefont {Wittek}, \citenamefont
  {Garcia-Gracia}, \citenamefont {El-Ganainy}, \citenamefont
  {Christodoulides},\ and\ \citenamefont {Khajavikhan}}]{Hodaei2017Aug}%
  \BibitemOpen
  \bibfield  {author} {\bibinfo {author} {\bibfnamefont {H.}~\bibnamefont
  {Hodaei}}, \bibinfo {author} {\bibfnamefont {A.~U.}\ \bibnamefont {Hassan}},
  \bibinfo {author} {\bibfnamefont {S.}~\bibnamefont {Wittek}}, \bibinfo
  {author} {\bibfnamefont {H.}~\bibnamefont {Garcia-Gracia}}, \bibinfo {author}
  {\bibfnamefont {R.}~\bibnamefont {El-Ganainy}}, \bibinfo {author}
  {\bibfnamefont {D.~N.}\ \bibnamefont {Christodoulides}},\ and\ \bibinfo
  {author} {\bibfnamefont {M.}~\bibnamefont {Khajavikhan}},\ }\href
  {https://doi.org/10.1038/nature23280} {\bibfield  {journal} {\bibinfo
  {journal} {Nature}\ }\textbf {\bibinfo {volume} {548}},\ \bibinfo {pages}
  {187} (\bibinfo {year} {2017})}\BibitemShut {NoStop}%
\bibitem [{\citenamefont {Song}\ \emph {et~al.}(2021)\citenamefont {Song},
  \citenamefont {Odeh}, \citenamefont {Z{\'u}{\~n}iga-P{\'e}rez}, \citenamefont
  {Kant{\'e}},\ and\ \citenamefont {Genevet}}]{song2021plasmonic}%
  \BibitemOpen
  \bibfield  {author} {\bibinfo {author} {\bibfnamefont {Q.}~\bibnamefont
  {Song}}, \bibinfo {author} {\bibfnamefont {M.}~\bibnamefont {Odeh}}, \bibinfo
  {author} {\bibfnamefont {J.}~\bibnamefont {Z{\'u}{\~n}iga-P{\'e}rez}},
  \bibinfo {author} {\bibfnamefont {B.}~\bibnamefont {Kant{\'e}}},\ and\
  \bibinfo {author} {\bibfnamefont {P.}~\bibnamefont {Genevet}},\ }\href@noop
  {} {\bibfield  {journal} {\bibinfo  {journal} {Science}\ }\textbf {\bibinfo
  {volume} {373}},\ \bibinfo {pages} {1133} (\bibinfo {year}
  {2021})}\BibitemShut {NoStop}%
\bibitem [{\citenamefont {Xu}\ \emph {et~al.}(2023)\citenamefont {Xu},
  \citenamefont {Li}, \citenamefont {Jeong}, \citenamefont {Kim}, \citenamefont
  {Kim}, \citenamefont {Rho},\ and\ \citenamefont {Liu}}]{xu2023subwavelength}%
  \BibitemOpen
  \bibfield  {author} {\bibinfo {author} {\bibfnamefont {Y.}~\bibnamefont
  {Xu}}, \bibinfo {author} {\bibfnamefont {L.}~\bibnamefont {Li}}, \bibinfo
  {author} {\bibfnamefont {H.}~\bibnamefont {Jeong}}, \bibinfo {author}
  {\bibfnamefont {S.}~\bibnamefont {Kim}}, \bibinfo {author} {\bibfnamefont
  {I.}~\bibnamefont {Kim}}, \bibinfo {author} {\bibfnamefont {J.}~\bibnamefont
  {Rho}},\ and\ \bibinfo {author} {\bibfnamefont {Y.}~\bibnamefont {Liu}},\
  }\href@noop {} {\bibfield  {journal} {\bibinfo  {journal} {Science advances}\
  }\textbf {\bibinfo {volume} {9}},\ \bibinfo {pages} {eadf3510} (\bibinfo
  {year} {2023})}\BibitemShut {NoStop}%
\bibitem [{\citenamefont {Park}\ \emph {et~al.}(2020)\citenamefont {Park},
  \citenamefont {Ndao}, \citenamefont {Cai}, \citenamefont {Hsu}, \citenamefont
  {Kodigala}, \citenamefont {Lepetit}, \citenamefont {Lo},\ and\ \citenamefont
  {Kant{\'e}}}]{park2020symmetry}%
  \BibitemOpen
  \bibfield  {author} {\bibinfo {author} {\bibfnamefont {J.-H.}\ \bibnamefont
  {Park}}, \bibinfo {author} {\bibfnamefont {A.}~\bibnamefont {Ndao}}, \bibinfo
  {author} {\bibfnamefont {W.}~\bibnamefont {Cai}}, \bibinfo {author}
  {\bibfnamefont {L.}~\bibnamefont {Hsu}}, \bibinfo {author} {\bibfnamefont
  {A.}~\bibnamefont {Kodigala}}, \bibinfo {author} {\bibfnamefont
  {T.}~\bibnamefont {Lepetit}}, \bibinfo {author} {\bibfnamefont {Y.-H.}\
  \bibnamefont {Lo}},\ and\ \bibinfo {author} {\bibfnamefont {B.}~\bibnamefont
  {Kant{\'e}}},\ }\href@noop {} {\bibfield  {journal} {\bibinfo  {journal}
  {Nature Physics}\ }\textbf {\bibinfo {volume} {16}},\ \bibinfo {pages} {462}
  (\bibinfo {year} {2020})}\BibitemShut {NoStop}%
\bibitem [{\citenamefont {Sang}\ \emph {et~al.}(2021)\citenamefont {Sang},
  \citenamefont {Wang}, \citenamefont {Raja}, \citenamefont {Cheng},
  \citenamefont {Huang}, \citenamefont {Chen}, \citenamefont {Zhang},
  \citenamefont {Ahn}, \citenamefont {Shih}, \citenamefont {Lee} \emph
  {et~al.}}]{sang2021tuning}%
  \BibitemOpen
  \bibfield  {author} {\bibinfo {author} {\bibfnamefont {Y.}~\bibnamefont
  {Sang}}, \bibinfo {author} {\bibfnamefont {C.-Y.}\ \bibnamefont {Wang}},
  \bibinfo {author} {\bibfnamefont {S.~S.}\ \bibnamefont {Raja}}, \bibinfo
  {author} {\bibfnamefont {C.-W.}\ \bibnamefont {Cheng}}, \bibinfo {author}
  {\bibfnamefont {C.-T.}\ \bibnamefont {Huang}}, \bibinfo {author}
  {\bibfnamefont {C.-A.}\ \bibnamefont {Chen}}, \bibinfo {author}
  {\bibfnamefont {X.-Q.}\ \bibnamefont {Zhang}}, \bibinfo {author}
  {\bibfnamefont {H.}~\bibnamefont {Ahn}}, \bibinfo {author} {\bibfnamefont
  {C.-K.}\ \bibnamefont {Shih}}, \bibinfo {author} {\bibfnamefont {Y.-H.}\
  \bibnamefont {Lee}}, \emph {et~al.},\ }\href@noop {} {\bibfield  {journal}
  {\bibinfo  {journal} {Nano letters}\ }\textbf {\bibinfo {volume} {21}},\
  \bibinfo {pages} {2596} (\bibinfo {year} {2021})}\BibitemShut {NoStop}%
\bibitem [{\citenamefont {Netherwood}\ \emph {et~al.}(2023)\citenamefont
  {Netherwood}, \citenamefont {Riley},\ and\ \citenamefont
  {Muljarov}}]{netherwood2023exceptional}%
  \BibitemOpen
  \bibfield  {author} {\bibinfo {author} {\bibfnamefont {K.~S.}\ \bibnamefont
  {Netherwood}}, \bibinfo {author} {\bibfnamefont {H.}~\bibnamefont {Riley}},\
  and\ \bibinfo {author} {\bibfnamefont {E.~A.}\ \bibnamefont {Muljarov}},\
  }\href@noop {} {\bibfield  {journal} {\bibinfo  {journal} {arXiv preprint
  arXiv:2309.12536}\ } (\bibinfo {year} {2023})}\BibitemShut {NoStop}%
\bibitem [{\citenamefont {Guo}\ \emph {et~al.}(2009)\citenamefont {Guo},
  \citenamefont {Salamo}, \citenamefont {Duchesne}, \citenamefont {Morandotti},
  \citenamefont {Volatier-Ravat}, \citenamefont {Aimez}, \citenamefont
  {Siviloglou},\ and\ \citenamefont {Christodoulides}}]{guo2009observation}%
  \BibitemOpen
  \bibfield  {author} {\bibinfo {author} {\bibfnamefont {A.}~\bibnamefont
  {Guo}}, \bibinfo {author} {\bibfnamefont {G.}~\bibnamefont {Salamo}},
  \bibinfo {author} {\bibfnamefont {D.}~\bibnamefont {Duchesne}}, \bibinfo
  {author} {\bibfnamefont {R.}~\bibnamefont {Morandotti}}, \bibinfo {author}
  {\bibfnamefont {M.}~\bibnamefont {Volatier-Ravat}}, \bibinfo {author}
  {\bibfnamefont {V.}~\bibnamefont {Aimez}}, \bibinfo {author} {\bibfnamefont
  {G.}~\bibnamefont {Siviloglou}},\ and\ \bibinfo {author} {\bibfnamefont
  {D.}~\bibnamefont {Christodoulides}},\ }\href@noop {} {\bibfield  {journal}
  {\bibinfo  {journal} {Physical review letters}\ }\textbf {\bibinfo {volume}
  {103}},\ \bibinfo {pages} {093902} (\bibinfo {year} {2009})}\BibitemShut
  {NoStop}%
\bibitem [{\citenamefont {Valero}\ \emph {et~al.}(2024)\citenamefont {Valero},
  \citenamefont {Bobrovs}, \citenamefont {Weiss}, \citenamefont {Gao},
  \citenamefont {Shalin},\ and\ \citenamefont
  {Kivshar}}]{valero2024bianisotropic}%
  \BibitemOpen
  \bibfield  {author} {\bibinfo {author} {\bibfnamefont {A.~C.}\ \bibnamefont
  {Valero}}, \bibinfo {author} {\bibfnamefont {V.}~\bibnamefont {Bobrovs}},
  \bibinfo {author} {\bibfnamefont {T.}~\bibnamefont {Weiss}}, \bibinfo
  {author} {\bibfnamefont {L.}~\bibnamefont {Gao}}, \bibinfo {author}
  {\bibfnamefont {A.~S.}\ \bibnamefont {Shalin}},\ and\ \bibinfo {author}
  {\bibfnamefont {Y.}~\bibnamefont {Kivshar}},\ }\href@noop {} {\bibfield
  {journal} {\bibinfo  {journal} {Physical Review Research}\ }\textbf {\bibinfo
  {volume} {6}},\ \bibinfo {pages} {013053} (\bibinfo {year}
  {2024})}\BibitemShut {NoStop}%
\bibitem [{\citenamefont {Hsu}\ \emph {et~al.}(2016)\citenamefont {Hsu},
  \citenamefont {Zhen}, \citenamefont {Stone}, \citenamefont {Joannopoulos},\
  and\ \citenamefont {Solja{\v{c}}i{\'c}}}]{hsu2016bound}%
  \BibitemOpen
  \bibfield  {author} {\bibinfo {author} {\bibfnamefont {C.~W.}\ \bibnamefont
  {Hsu}}, \bibinfo {author} {\bibfnamefont {B.}~\bibnamefont {Zhen}}, \bibinfo
  {author} {\bibfnamefont {A.~D.}\ \bibnamefont {Stone}}, \bibinfo {author}
  {\bibfnamefont {J.~D.}\ \bibnamefont {Joannopoulos}},\ and\ \bibinfo {author}
  {\bibfnamefont {M.}~\bibnamefont {Solja{\v{c}}i{\'c}}},\ }\href@noop {}
  {\bibfield  {journal} {\bibinfo  {journal} {Nature Reviews Materials}\
  }\textbf {\bibinfo {volume} {1}},\ \bibinfo {pages} {1} (\bibinfo {year}
  {2016})}\BibitemShut {NoStop}%
\bibitem [{\citenamefont {Koshelev}\ \emph {et~al.}(2023)\citenamefont
  {Koshelev}, \citenamefont {Sadrieva}, \citenamefont {Shcherbakov},
  \citenamefont {Kivshar},\ and\ \citenamefont {Bogdanov}}]{Koshelev2023May}%
  \BibitemOpen
  \bibfield  {author} {\bibinfo {author} {\bibfnamefont {K.~L.}\ \bibnamefont
  {Koshelev}}, \bibinfo {author} {\bibfnamefont {Z.~F.}\ \bibnamefont
  {Sadrieva}}, \bibinfo {author} {\bibfnamefont {A.~A.}\ \bibnamefont
  {Shcherbakov}}, \bibinfo {author} {\bibfnamefont {{\relax Yu}.~S.}\
  \bibnamefont {Kivshar}},\ and\ \bibinfo {author} {\bibfnamefont {A.~A.}\
  \bibnamefont {Bogdanov}},\ }\href {https://ufn.ru/en/articles/2023/5/c}
  {\bibfield  {journal} {\bibinfo  {journal} {Phys.-Usp.}\ }\textbf {\bibinfo
  {volume} {66}},\ \bibinfo {pages} {494} (\bibinfo {year} {2023})}\BibitemShut
  {NoStop}%
\bibitem [{\citenamefont {Zhen}\ \emph {et~al.}(2014)\citenamefont {Zhen},
  \citenamefont {Hsu}, \citenamefont {Lu}, \citenamefont {Stone},\ and\
  \citenamefont {Solja{\v{c}}i{\'c}}}]{zhen2014topological}%
  \BibitemOpen
  \bibfield  {author} {\bibinfo {author} {\bibfnamefont {B.}~\bibnamefont
  {Zhen}}, \bibinfo {author} {\bibfnamefont {C.~W.}\ \bibnamefont {Hsu}},
  \bibinfo {author} {\bibfnamefont {L.}~\bibnamefont {Lu}}, \bibinfo {author}
  {\bibfnamefont {A.~D.}\ \bibnamefont {Stone}},\ and\ \bibinfo {author}
  {\bibfnamefont {M.}~\bibnamefont {Solja{\v{c}}i{\'c}}},\ }\href@noop {}
  {\bibfield  {journal} {\bibinfo  {journal} {Physical review letters}\
  }\textbf {\bibinfo {volume} {113}},\ \bibinfo {pages} {257401} (\bibinfo
  {year} {2014})}\BibitemShut {NoStop}%
\bibitem [{\citenamefont {Von~Neumann}\ and\ \citenamefont
  {Wigner}(1929)}]{von1929some}%
  \BibitemOpen
  \bibfield  {author} {\bibinfo {author} {\bibfnamefont {J.}~\bibnamefont
  {Von~Neumann}}\ and\ \bibinfo {author} {\bibfnamefont {E.}~\bibnamefont
  {Wigner}},\ }\href@noop {} {\bibfield  {journal} {\bibinfo  {journal} {Phys.
  Z}\ }\textbf {\bibinfo {volume} {30}},\ \bibinfo {pages} {465} (\bibinfo
  {year} {1929})}\BibitemShut {NoStop}%
\bibitem [{\citenamefont {Stillinger}\ and\ \citenamefont
  {Herrick}(1975)}]{stillinger1975bound}%
  \BibitemOpen
  \bibfield  {author} {\bibinfo {author} {\bibfnamefont {F.~H.}\ \bibnamefont
  {Stillinger}}\ and\ \bibinfo {author} {\bibfnamefont {D.~R.}\ \bibnamefont
  {Herrick}},\ }\href@noop {} {\bibfield  {journal} {\bibinfo  {journal}
  {Physical Review A}\ }\textbf {\bibinfo {volume} {11}},\ \bibinfo {pages}
  {446} (\bibinfo {year} {1975})}\BibitemShut {NoStop}%
\bibitem [{\citenamefont {Azzam}\ and\ \citenamefont
  {Kildishev}(2021)}]{azzam2021photonic}%
  \BibitemOpen
  \bibfield  {author} {\bibinfo {author} {\bibfnamefont {S.~I.}\ \bibnamefont
  {Azzam}}\ and\ \bibinfo {author} {\bibfnamefont {A.~V.}\ \bibnamefont
  {Kildishev}},\ }\href@noop {} {\bibfield  {journal} {\bibinfo  {journal}
  {Advanced Optical Materials}\ }\textbf {\bibinfo {volume} {9}},\ \bibinfo
  {pages} {2001469} (\bibinfo {year} {2021})}\BibitemShut {NoStop}%
\bibitem [{\citenamefont {Plotnik}\ \emph {et~al.}(2011)\citenamefont
  {Plotnik}, \citenamefont {Peleg}, \citenamefont {Dreisow}, \citenamefont
  {Heinrich}, \citenamefont {Nolte}, \citenamefont {Szameit},\ and\
  \citenamefont {Segev}}]{plotnik2011experimental}%
  \BibitemOpen
  \bibfield  {author} {\bibinfo {author} {\bibfnamefont {Y.}~\bibnamefont
  {Plotnik}}, \bibinfo {author} {\bibfnamefont {O.}~\bibnamefont {Peleg}},
  \bibinfo {author} {\bibfnamefont {F.}~\bibnamefont {Dreisow}}, \bibinfo
  {author} {\bibfnamefont {M.}~\bibnamefont {Heinrich}}, \bibinfo {author}
  {\bibfnamefont {S.}~\bibnamefont {Nolte}}, \bibinfo {author} {\bibfnamefont
  {A.}~\bibnamefont {Szameit}},\ and\ \bibinfo {author} {\bibfnamefont
  {M.}~\bibnamefont {Segev}},\ }\href@noop {} {\bibfield  {journal} {\bibinfo
  {journal} {Physical review letters}\ }\textbf {\bibinfo {volume} {107}},\
  \bibinfo {pages} {183901} (\bibinfo {year} {2011})}\BibitemShut {NoStop}%
\bibitem [{\citenamefont {Liang}\ \emph {et~al.}(2020)\citenamefont {Liang},
  \citenamefont {Koshelev}, \citenamefont {Zhang}, \citenamefont {Lin},
  \citenamefont {Lin}, \citenamefont {Wu}, \citenamefont {Jia},\ and\
  \citenamefont {Kivshar}}]{Liang2020Sep}%
  \BibitemOpen
  \bibfield  {author} {\bibinfo {author} {\bibfnamefont {Y.}~\bibnamefont
  {Liang}}, \bibinfo {author} {\bibfnamefont {K.}~\bibnamefont {Koshelev}},
  \bibinfo {author} {\bibfnamefont {F.}~\bibnamefont {Zhang}}, \bibinfo
  {author} {\bibfnamefont {H.}~\bibnamefont {Lin}}, \bibinfo {author}
  {\bibfnamefont {S.}~\bibnamefont {Lin}}, \bibinfo {author} {\bibfnamefont
  {J.}~\bibnamefont {Wu}}, \bibinfo {author} {\bibfnamefont {B.}~\bibnamefont
  {Jia}},\ and\ \bibinfo {author} {\bibfnamefont {Y.}~\bibnamefont {Kivshar}},\
  }\href {https://doi.org/10.1021/acs.nanolett.0c01752} {\bibfield  {journal}
  {\bibinfo  {journal} {Nano Lett.}\ }\textbf {\bibinfo {volume} {20}},\
  \bibinfo {pages} {6351} (\bibinfo {year} {2020})}\BibitemShut {NoStop}%
\bibitem [{\citenamefont {Azzam}\ \emph
  {et~al.}(2018{\natexlab{a}})\citenamefont {Azzam}, \citenamefont {Shalaev},
  \citenamefont {Boltasseva},\ and\ \citenamefont {Kildishev}}]{Azzam2018Dec}%
  \BibitemOpen
  \bibfield  {author} {\bibinfo {author} {\bibfnamefont {S.~I.}\ \bibnamefont
  {Azzam}}, \bibinfo {author} {\bibfnamefont {V.~M.}\ \bibnamefont {Shalaev}},
  \bibinfo {author} {\bibfnamefont {A.}~\bibnamefont {Boltasseva}},\ and\
  \bibinfo {author} {\bibfnamefont {A.~V.}\ \bibnamefont {Kildishev}},\ }\href
  {https://doi.org/10.1103/PhysRevLett.121.253901} {\bibfield  {journal}
  {\bibinfo  {journal} {Phys. Rev. Lett.}\ }\textbf {\bibinfo {volume} {121}},\
  \bibinfo {pages} {253901} (\bibinfo {year} {2018}{\natexlab{a}})}\BibitemShut
  {NoStop}%
\bibitem [{\citenamefont {Neale}\ and\ \citenamefont
  {Muljarov}(2021)}]{neale2021accidental}%
  \BibitemOpen
  \bibfield  {author} {\bibinfo {author} {\bibfnamefont {S.}~\bibnamefont
  {Neale}}\ and\ \bibinfo {author} {\bibfnamefont {E.~A.}\ \bibnamefont
  {Muljarov}},\ }\href@noop {} {\bibfield  {journal} {\bibinfo  {journal}
  {Physical Review B}\ }\textbf {\bibinfo {volume} {103}},\ \bibinfo {pages}
  {155112} (\bibinfo {year} {2021})}\BibitemShut {NoStop}%
\bibitem [{\citenamefont {Koshelev}\ \emph {et~al.}(2018)\citenamefont
  {Koshelev}, \citenamefont {Lepeshov}, \citenamefont {Liu}, \citenamefont
  {Bogdanov},\ and\ \citenamefont {Kivshar}}]{koshelev2018asymmetric}%
  \BibitemOpen
  \bibfield  {author} {\bibinfo {author} {\bibfnamefont {K.}~\bibnamefont
  {Koshelev}}, \bibinfo {author} {\bibfnamefont {S.}~\bibnamefont {Lepeshov}},
  \bibinfo {author} {\bibfnamefont {M.}~\bibnamefont {Liu}}, \bibinfo {author}
  {\bibfnamefont {A.}~\bibnamefont {Bogdanov}},\ and\ \bibinfo {author}
  {\bibfnamefont {Y.}~\bibnamefont {Kivshar}},\ }\href
  {https://doi.org/10.1103/PhysRevLett.121.193903} {\bibfield  {journal}
  {\bibinfo  {journal} {Phys. Rev. Lett.}\ }\textbf {\bibinfo {volume} {121}},\
  \bibinfo {pages} {193903} (\bibinfo {year} {2018})}\BibitemShut {NoStop}%
\bibitem [{\citenamefont {Yesilkoy}\ \emph {et~al.}(2019)\citenamefont
  {Yesilkoy}, \citenamefont {Arvelo}, \citenamefont {Jahani}, \citenamefont
  {Liu}, \citenamefont {Tittl}, \citenamefont {Cevher}, \citenamefont
  {Kivshar},\ and\ \citenamefont {Altug}}]{Yesilkoy2019Jun}%
  \BibitemOpen
  \bibfield  {author} {\bibinfo {author} {\bibfnamefont {F.}~\bibnamefont
  {Yesilkoy}}, \bibinfo {author} {\bibfnamefont {E.~R.}\ \bibnamefont
  {Arvelo}}, \bibinfo {author} {\bibfnamefont {Y.}~\bibnamefont {Jahani}},
  \bibinfo {author} {\bibfnamefont {M.}~\bibnamefont {Liu}}, \bibinfo {author}
  {\bibfnamefont {A.}~\bibnamefont {Tittl}}, \bibinfo {author} {\bibfnamefont
  {V.}~\bibnamefont {Cevher}}, \bibinfo {author} {\bibfnamefont
  {Y.}~\bibnamefont {Kivshar}},\ and\ \bibinfo {author} {\bibfnamefont
  {H.}~\bibnamefont {Altug}},\ }\href
  {https://doi.org/10.1038/s41566-019-0394-6} {\bibfield  {journal} {\bibinfo
  {journal} {Nat. Photonics}\ }\textbf {\bibinfo {volume} {13}},\ \bibinfo
  {pages} {390} (\bibinfo {year} {2019})}\BibitemShut {NoStop}%
\bibitem [{\citenamefont {Kravtsov}\ \emph {et~al.}(2020)\citenamefont
  {Kravtsov}, \citenamefont {Khestanova}, \citenamefont {Benimetskiy},
  \citenamefont {Ivanova}, \citenamefont {Samusev}, \citenamefont {Sinev},
  \citenamefont {Pidgayko}, \citenamefont {Mozharov}, \citenamefont {Mukhin},
  \citenamefont {Lozhkin}, \citenamefont {Kapitonov}, \citenamefont {Brichkin},
  \citenamefont {Kulakovskii}, \citenamefont {Shelykh}, \citenamefont
  {Tartakovskii}, \citenamefont {Walker}, \citenamefont {Skolnick},
  \citenamefont {Krizhanovskii},\ and\ \citenamefont
  {Iorsh}}]{Kravtsov2020Apr}%
  \BibitemOpen
  \bibfield  {author} {\bibinfo {author} {\bibfnamefont {V.}~\bibnamefont
  {Kravtsov}}, \bibinfo {author} {\bibfnamefont {E.}~\bibnamefont
  {Khestanova}}, \bibinfo {author} {\bibfnamefont {F.~A.}\ \bibnamefont
  {Benimetskiy}}, \bibinfo {author} {\bibfnamefont {T.}~\bibnamefont
  {Ivanova}}, \bibinfo {author} {\bibfnamefont {A.~K.}\ \bibnamefont
  {Samusev}}, \bibinfo {author} {\bibfnamefont {I.~S.}\ \bibnamefont {Sinev}},
  \bibinfo {author} {\bibfnamefont {D.}~\bibnamefont {Pidgayko}}, \bibinfo
  {author} {\bibfnamefont {A.~M.}\ \bibnamefont {Mozharov}}, \bibinfo {author}
  {\bibfnamefont {I.~S.}\ \bibnamefont {Mukhin}}, \bibinfo {author}
  {\bibfnamefont {M.~S.}\ \bibnamefont {Lozhkin}}, \bibinfo {author}
  {\bibfnamefont {Y.~V.}\ \bibnamefont {Kapitonov}}, \bibinfo {author}
  {\bibfnamefont {A.~S.}\ \bibnamefont {Brichkin}}, \bibinfo {author}
  {\bibfnamefont {V.~D.}\ \bibnamefont {Kulakovskii}}, \bibinfo {author}
  {\bibfnamefont {I.~A.}\ \bibnamefont {Shelykh}}, \bibinfo {author}
  {\bibfnamefont {A.~I.}\ \bibnamefont {Tartakovskii}}, \bibinfo {author}
  {\bibfnamefont {P.~M.}\ \bibnamefont {Walker}}, \bibinfo {author}
  {\bibfnamefont {M.~S.}\ \bibnamefont {Skolnick}}, \bibinfo {author}
  {\bibfnamefont {D.~N.}\ \bibnamefont {Krizhanovskii}},\ and\ \bibinfo
  {author} {\bibfnamefont {I.~V.}\ \bibnamefont {Iorsh}},\ }\href
  {https://doi.org/10.1038/s41377-020-0286-z} {\bibfield  {journal} {\bibinfo
  {journal} {Light Sci. Appl.}\ }\textbf {\bibinfo {volume} {9}},\ \bibinfo
  {pages} {1} (\bibinfo {year} {2020})}\BibitemShut {NoStop}%
\bibitem [{\citenamefont {Maggiolini}\ \emph {et~al.}(2023)\citenamefont
  {Maggiolini}, \citenamefont {Polimeno}, \citenamefont {Todisco},
  \citenamefont {Di~Renzo}, \citenamefont {Han}, \citenamefont {De~Giorgi},
  \citenamefont {Ardizzone}, \citenamefont {Schneider}, \citenamefont
  {Mastria}, \citenamefont {Cannavale}, \citenamefont {Pugliese}, \citenamefont
  {De~Marco}, \citenamefont {Rizzo}, \citenamefont {Maiorano}, \citenamefont
  {Gigli}, \citenamefont {Gerace}, \citenamefont {Sanvitto},\ and\
  \citenamefont {Ballarini}}]{Maggiolini2023Aug}%
  \BibitemOpen
  \bibfield  {author} {\bibinfo {author} {\bibfnamefont {E.}~\bibnamefont
  {Maggiolini}}, \bibinfo {author} {\bibfnamefont {L.}~\bibnamefont
  {Polimeno}}, \bibinfo {author} {\bibfnamefont {F.}~\bibnamefont {Todisco}},
  \bibinfo {author} {\bibfnamefont {A.}~\bibnamefont {Di~Renzo}}, \bibinfo
  {author} {\bibfnamefont {B.}~\bibnamefont {Han}}, \bibinfo {author}
  {\bibfnamefont {M.}~\bibnamefont {De~Giorgi}}, \bibinfo {author}
  {\bibfnamefont {V.}~\bibnamefont {Ardizzone}}, \bibinfo {author}
  {\bibfnamefont {C.}~\bibnamefont {Schneider}}, \bibinfo {author}
  {\bibfnamefont {R.}~\bibnamefont {Mastria}}, \bibinfo {author} {\bibfnamefont
  {A.}~\bibnamefont {Cannavale}}, \bibinfo {author} {\bibfnamefont
  {M.}~\bibnamefont {Pugliese}}, \bibinfo {author} {\bibfnamefont
  {L.}~\bibnamefont {De~Marco}}, \bibinfo {author} {\bibfnamefont
  {A.}~\bibnamefont {Rizzo}}, \bibinfo {author} {\bibfnamefont
  {V.}~\bibnamefont {Maiorano}}, \bibinfo {author} {\bibfnamefont
  {G.}~\bibnamefont {Gigli}}, \bibinfo {author} {\bibfnamefont
  {D.}~\bibnamefont {Gerace}}, \bibinfo {author} {\bibfnamefont
  {D.}~\bibnamefont {Sanvitto}},\ and\ \bibinfo {author} {\bibfnamefont
  {D.}~\bibnamefont {Ballarini}},\ }\href
  {https://doi.org/10.1038/s41563-023-01562-9} {\bibfield  {journal} {\bibinfo
  {journal} {Nat. Mater.}\ }\textbf {\bibinfo {volume} {22}},\ \bibinfo {pages}
  {964} (\bibinfo {year} {2023})}\BibitemShut {NoStop}%
\bibitem [{\citenamefont {Yang}\ \emph {et~al.}(2020)\citenamefont {Yang},
  \citenamefont {Wang}, \citenamefont {Rao}, \citenamefont {Gui}, \citenamefont
  {Yao}, \citenamefont {Lu},\ and\ \citenamefont
  {Hu}}]{yang2020unconventional}%
  \BibitemOpen
  \bibfield  {author} {\bibinfo {author} {\bibfnamefont {Y.}~\bibnamefont
  {Yang}}, \bibinfo {author} {\bibfnamefont {Y.-P.}\ \bibnamefont {Wang}},
  \bibinfo {author} {\bibfnamefont {J.}~\bibnamefont {Rao}}, \bibinfo {author}
  {\bibfnamefont {Y.}~\bibnamefont {Gui}}, \bibinfo {author} {\bibfnamefont
  {B.}~\bibnamefont {Yao}}, \bibinfo {author} {\bibfnamefont {W.}~\bibnamefont
  {Lu}},\ and\ \bibinfo {author} {\bibfnamefont {C.-M.}\ \bibnamefont {Hu}},\
  }\href@noop {} {\bibfield  {journal} {\bibinfo  {journal} {Physical Review
  Letters}\ }\textbf {\bibinfo {volume} {125}},\ \bibinfo {pages} {147202}
  (\bibinfo {year} {2020})}\BibitemShut {NoStop}%
\bibitem [{\citenamefont {Deng}\ \emph {et~al.}(2022)\citenamefont {Deng},
  \citenamefont {Li}, \citenamefont {Li}, \citenamefont {Li},\ and\
  \citenamefont {Al{\`u}}}]{deng2022extreme}%
  \BibitemOpen
  \bibfield  {author} {\bibinfo {author} {\bibfnamefont {Z.-L.}\ \bibnamefont
  {Deng}}, \bibinfo {author} {\bibfnamefont {F.-J.}\ \bibnamefont {Li}},
  \bibinfo {author} {\bibfnamefont {H.}~\bibnamefont {Li}}, \bibinfo {author}
  {\bibfnamefont {X.}~\bibnamefont {Li}},\ and\ \bibinfo {author}
  {\bibfnamefont {A.}~\bibnamefont {Al{\`u}}},\ }\href@noop {} {\bibfield
  {journal} {\bibinfo  {journal} {Laser \& Photonics Reviews}\ }\textbf
  {\bibinfo {volume} {16}},\ \bibinfo {pages} {2100617} (\bibinfo {year}
  {2022})}\BibitemShut {NoStop}%
\bibitem [{\citenamefont {Sakotic}\ \emph {et~al.}(2023)\citenamefont
  {Sakotic}, \citenamefont {Stankovic}, \citenamefont {Bengin}, \citenamefont
  {Krasnok}, \citenamefont {Al{\'u}},\ and\ \citenamefont
  {Jankovic}}]{sakotic2023non}%
  \BibitemOpen
  \bibfield  {author} {\bibinfo {author} {\bibfnamefont {Z.}~\bibnamefont
  {Sakotic}}, \bibinfo {author} {\bibfnamefont {P.}~\bibnamefont {Stankovic}},
  \bibinfo {author} {\bibfnamefont {V.}~\bibnamefont {Bengin}}, \bibinfo
  {author} {\bibfnamefont {A.}~\bibnamefont {Krasnok}}, \bibinfo {author}
  {\bibfnamefont {A.}~\bibnamefont {Al{\'u}}},\ and\ \bibinfo {author}
  {\bibfnamefont {N.}~\bibnamefont {Jankovic}},\ }\href@noop {} {\bibfield
  {journal} {\bibinfo  {journal} {Laser \& Photonics Reviews}\ ,\ \bibinfo
  {pages} {2200308}} (\bibinfo {year} {2023})}\BibitemShut {NoStop}%
\bibitem [{\citenamefont {Qin}\ \emph {et~al.}(2022)\citenamefont {Qin},
  \citenamefont {Shi},\ and\ \citenamefont {Ou}}]{qin2022exceptional}%
  \BibitemOpen
  \bibfield  {author} {\bibinfo {author} {\bibfnamefont {H.}~\bibnamefont
  {Qin}}, \bibinfo {author} {\bibfnamefont {X.}~\bibnamefont {Shi}},\ and\
  \bibinfo {author} {\bibfnamefont {H.}~\bibnamefont {Ou}},\ }\href@noop {}
  {\bibfield  {journal} {\bibinfo  {journal} {Nanophotonics}\ }\textbf
  {\bibinfo {volume} {11}},\ \bibinfo {pages} {4909} (\bibinfo {year}
  {2022})}\BibitemShut {NoStop}%
\bibitem [{\citenamefont {Kikkawa}\ \emph {et~al.}(2020)\citenamefont
  {Kikkawa}, \citenamefont {Nishida},\ and\ \citenamefont
  {Kadoya}}]{kikkawa2020bound}%
  \BibitemOpen
  \bibfield  {author} {\bibinfo {author} {\bibfnamefont {R.}~\bibnamefont
  {Kikkawa}}, \bibinfo {author} {\bibfnamefont {M.}~\bibnamefont {Nishida}},\
  and\ \bibinfo {author} {\bibfnamefont {Y.}~\bibnamefont {Kadoya}},\
  }\href@noop {} {\bibfield  {journal} {\bibinfo  {journal} {New Journal of
  Physics}\ }\textbf {\bibinfo {volume} {22}},\ \bibinfo {pages} {073029}
  (\bibinfo {year} {2020})}\BibitemShut {NoStop}%
\bibitem [{\citenamefont {Aigner}\ \emph {et~al.}(2022)\citenamefont {Aigner},
  \citenamefont {Tittl}, \citenamefont {Wang}, \citenamefont {Weber},
  \citenamefont {Kivshar}, \citenamefont {Maier},\ and\ \citenamefont
  {Ren}}]{aigner2022plasmonic}%
  \BibitemOpen
  \bibfield  {author} {\bibinfo {author} {\bibfnamefont {A.}~\bibnamefont
  {Aigner}}, \bibinfo {author} {\bibfnamefont {A.}~\bibnamefont {Tittl}},
  \bibinfo {author} {\bibfnamefont {J.}~\bibnamefont {Wang}}, \bibinfo {author}
  {\bibfnamefont {T.}~\bibnamefont {Weber}}, \bibinfo {author} {\bibfnamefont
  {Y.}~\bibnamefont {Kivshar}}, \bibinfo {author} {\bibfnamefont {S.~A.}\
  \bibnamefont {Maier}},\ and\ \bibinfo {author} {\bibfnamefont
  {H.}~\bibnamefont {Ren}},\ }\href@noop {} {\bibfield  {journal} {\bibinfo
  {journal} {Science advances}\ }\textbf {\bibinfo {volume} {8}},\ \bibinfo
  {pages} {eadd4816} (\bibinfo {year} {2022})}\BibitemShut {NoStop}%
\bibitem [{\citenamefont {Tao}\ \emph {et~al.}(2023)\citenamefont {Tao},
  \citenamefont {Fang}, \citenamefont {Pan}, \citenamefont {Zhang},
  \citenamefont {Jin}, \citenamefont {Hong},\ and\ \citenamefont
  {Shu}}]{tao2023tunable}%
  \BibitemOpen
  \bibfield  {author} {\bibinfo {author} {\bibfnamefont {Y.}~\bibnamefont
  {Tao}}, \bibinfo {author} {\bibfnamefont {B.}~\bibnamefont {Fang}}, \bibinfo
  {author} {\bibfnamefont {G.-M.}\ \bibnamefont {Pan}}, \bibinfo {author}
  {\bibfnamefont {D.-Q.}\ \bibnamefont {Zhang}}, \bibinfo {author}
  {\bibfnamefont {Z.-W.}\ \bibnamefont {Jin}}, \bibinfo {author} {\bibfnamefont
  {Z.}~\bibnamefont {Hong}},\ and\ \bibinfo {author} {\bibfnamefont {F.-Z.}\
  \bibnamefont {Shu}},\ }\href@noop {} {\bibfield  {journal} {\bibinfo
  {journal} {ACS Applied Optical Materials}\ }\textbf {\bibinfo {volume} {1}},\
  \bibinfo {pages} {1452} (\bibinfo {year} {2023})}\BibitemShut {NoStop}%
\bibitem [{\citenamefont {Both}\ and\ \citenamefont
  {Weiss}(2021)}]{both2021resonant}%
  \BibitemOpen
  \bibfield  {author} {\bibinfo {author} {\bibfnamefont {S.}~\bibnamefont
  {Both}}\ and\ \bibinfo {author} {\bibfnamefont {T.}~\bibnamefont {Weiss}},\
  }\href@noop {} {\bibfield  {journal} {\bibinfo  {journal} {Semiconductor
  Science and Technology}\ }\textbf {\bibinfo {volume} {37}},\ \bibinfo {pages}
  {013002} (\bibinfo {year} {2021})}\BibitemShut {NoStop}%
\bibitem [{\citenamefont {Kristensen}\ \emph {et~al.}(2020)\citenamefont
  {Kristensen}, \citenamefont {Herrmann}, \citenamefont {Intravaia},\ and\
  \citenamefont {Busch}}]{kristensen2020modeling}%
  \BibitemOpen
  \bibfield  {author} {\bibinfo {author} {\bibfnamefont {P.~T.}\ \bibnamefont
  {Kristensen}}, \bibinfo {author} {\bibfnamefont {K.}~\bibnamefont
  {Herrmann}}, \bibinfo {author} {\bibfnamefont {F.}~\bibnamefont
  {Intravaia}},\ and\ \bibinfo {author} {\bibfnamefont {K.}~\bibnamefont
  {Busch}},\ }\href@noop {} {\bibfield  {journal} {\bibinfo  {journal}
  {Advances in Optics and Photonics}\ }\textbf {\bibinfo {volume} {12}},\
  \bibinfo {pages} {612} (\bibinfo {year} {2020})}\BibitemShut {NoStop}%
\bibitem [{\citenamefont {Lalanne}\ \emph {et~al.}(2018)\citenamefont
  {Lalanne}, \citenamefont {Yan}, \citenamefont {Vynck}, \citenamefont
  {Sauvan},\ and\ \citenamefont {Hugonin}}]{lalanne2018light}%
  \BibitemOpen
  \bibfield  {author} {\bibinfo {author} {\bibfnamefont {P.}~\bibnamefont
  {Lalanne}}, \bibinfo {author} {\bibfnamefont {W.}~\bibnamefont {Yan}},
  \bibinfo {author} {\bibfnamefont {K.}~\bibnamefont {Vynck}}, \bibinfo
  {author} {\bibfnamefont {C.}~\bibnamefont {Sauvan}},\ and\ \bibinfo {author}
  {\bibfnamefont {J.-P.}\ \bibnamefont {Hugonin}},\ }\href@noop {} {\bibfield
  {journal} {\bibinfo  {journal} {Laser \& Photonics Reviews}\ }\textbf
  {\bibinfo {volume} {12}},\ \bibinfo {pages} {1700113} (\bibinfo {year}
  {2018})}\BibitemShut {NoStop}%
\bibitem [{\citenamefont {Sadreev}(2021)}]{sadreev2021interference}%
  \BibitemOpen
  \bibfield  {author} {\bibinfo {author} {\bibfnamefont {A.~F.}\ \bibnamefont
  {Sadreev}},\ }\href@noop {} {\bibfield  {journal} {\bibinfo  {journal}
  {Reports on Progress in Physics}\ }\textbf {\bibinfo {volume} {84}},\
  \bibinfo {pages} {055901} (\bibinfo {year} {2021})}\BibitemShut {NoStop}%
\bibitem [{\citenamefont {Shabanov}(2009)}]{shabanov2009resonant}%
  \BibitemOpen
  \bibfield  {author} {\bibinfo {author} {\bibfnamefont {S.~V.}\ \bibnamefont
  {Shabanov}},\ }\href@noop {} {\bibfield  {journal} {\bibinfo  {journal}
  {International Journal of Modern Physics B}\ }\textbf {\bibinfo {volume}
  {23}},\ \bibinfo {pages} {5191} (\bibinfo {year} {2009})}\BibitemShut
  {NoStop}%
\bibitem [{\citenamefont {Doiron}\ \emph {et~al.}(2022)\citenamefont {Doiron},
  \citenamefont {Brener},\ and\ \citenamefont {Cerjan}}]{doiron2022realizing}%
  \BibitemOpen
  \bibfield  {author} {\bibinfo {author} {\bibfnamefont {C.~F.}\ \bibnamefont
  {Doiron}}, \bibinfo {author} {\bibfnamefont {I.}~\bibnamefont {Brener}},\
  and\ \bibinfo {author} {\bibfnamefont {A.}~\bibnamefont {Cerjan}},\
  }\href@noop {} {\bibfield  {journal} {\bibinfo  {journal} {Nature
  Communications}\ }\textbf {\bibinfo {volume} {13}},\ \bibinfo {pages} {7534}
  (\bibinfo {year} {2022})}\BibitemShut {NoStop}%
\bibitem [{\citenamefont {Teller}(1937)}]{teller1937crossing}%
  \BibitemOpen
  \bibfield  {author} {\bibinfo {author} {\bibfnamefont {E.}~\bibnamefont
  {Teller}},\ }\href@noop {} {\bibfield  {journal} {\bibinfo  {journal}
  {Journal of Physical Chemistry}\ }\textbf {\bibinfo {volume} {41}},\ \bibinfo
  {pages} {109} (\bibinfo {year} {1937})}\BibitemShut {NoStop}%
\bibitem [{Sup()}]{Suppl_Mat_1}%
  \BibitemOpen
  \href@noop {} {}\bibinfo {note} {See the Supplemental Material for (i)
  non-existence of the EP-BIC for conventional lossless systems, (ii) coupled
  mode theory of resonant states, (iii) simplified effective Hamiltonian,(iv)
  assymptotics of an EP-BIC with the introduction of radiative loss,(v)
  separation of losses into absorptive and radiative, (vi) reflection spectra
  with different off-center holes, (vii) EP-BIC dispersion in reciprocal
  space.}\BibitemShut {Stop}%
\bibitem [{\citenamefont {Tao}\ \emph {et~al.}(2020)\citenamefont {Tao},
  \citenamefont {Zhu}, \citenamefont {Zhong},\ and\ \citenamefont
  {Liu}}]{tao2020coupling}%
  \BibitemOpen
  \bibfield  {author} {\bibinfo {author} {\bibfnamefont {C.}~\bibnamefont
  {Tao}}, \bibinfo {author} {\bibfnamefont {J.}~\bibnamefont {Zhu}}, \bibinfo
  {author} {\bibfnamefont {Y.}~\bibnamefont {Zhong}},\ and\ \bibinfo {author}
  {\bibfnamefont {H.}~\bibnamefont {Liu}},\ }\href@noop {} {\bibfield
  {journal} {\bibinfo  {journal} {Physical Review B}\ }\textbf {\bibinfo
  {volume} {102}},\ \bibinfo {pages} {045430} (\bibinfo {year}
  {2020})}\BibitemShut {NoStop}%
\bibitem [{\citenamefont {Zhou}\ \emph {et~al.}(2018)\citenamefont {Zhou},
  \citenamefont {Peng}, \citenamefont {Yoon}, \citenamefont {Hsu},
  \citenamefont {Nelson}, \citenamefont {Fu}, \citenamefont {Joannopoulos},
  \citenamefont {Solja{\v{c}}i{\'c}},\ and\ \citenamefont
  {Zhen}}]{zhou2018observation}%
  \BibitemOpen
  \bibfield  {author} {\bibinfo {author} {\bibfnamefont {H.}~\bibnamefont
  {Zhou}}, \bibinfo {author} {\bibfnamefont {C.}~\bibnamefont {Peng}}, \bibinfo
  {author} {\bibfnamefont {Y.}~\bibnamefont {Yoon}}, \bibinfo {author}
  {\bibfnamefont {C.~W.}\ \bibnamefont {Hsu}}, \bibinfo {author} {\bibfnamefont
  {K.~A.}\ \bibnamefont {Nelson}}, \bibinfo {author} {\bibfnamefont
  {L.}~\bibnamefont {Fu}}, \bibinfo {author} {\bibfnamefont {J.~D.}\
  \bibnamefont {Joannopoulos}}, \bibinfo {author} {\bibfnamefont
  {M.}~\bibnamefont {Solja{\v{c}}i{\'c}}},\ and\ \bibinfo {author}
  {\bibfnamefont {B.}~\bibnamefont {Zhen}},\ }\href@noop {} {\bibfield
  {journal} {\bibinfo  {journal} {Science}\ }\textbf {\bibinfo {volume}
  {359}},\ \bibinfo {pages} {1009} (\bibinfo {year} {2018})}\BibitemShut
  {NoStop}%
\bibitem [{\citenamefont {Weiss}\ \emph {et~al.}(2016)\citenamefont {Weiss},
  \citenamefont {Mesch}, \citenamefont {Sch{\"a}ferling}, \citenamefont
  {Giessen}, \citenamefont {Langbein},\ and\ \citenamefont
  {Muljarov}}]{weiss2016dark}%
  \BibitemOpen
  \bibfield  {author} {\bibinfo {author} {\bibfnamefont {T.}~\bibnamefont
  {Weiss}}, \bibinfo {author} {\bibfnamefont {M.}~\bibnamefont {Mesch}},
  \bibinfo {author} {\bibfnamefont {M.}~\bibnamefont {Sch{\"a}ferling}},
  \bibinfo {author} {\bibfnamefont {H.}~\bibnamefont {Giessen}}, \bibinfo
  {author} {\bibfnamefont {W.}~\bibnamefont {Langbein}},\ and\ \bibinfo
  {author} {\bibfnamefont {E.~A.}\ \bibnamefont {Muljarov}},\ }\href@noop {}
  {\bibfield  {journal} {\bibinfo  {journal} {Physical review letters}\
  }\textbf {\bibinfo {volume} {116}},\ \bibinfo {pages} {237401} (\bibinfo
  {year} {2016})}\BibitemShut {NoStop}%
\bibitem [{\citenamefont {Wang}\ \emph {et~al.}(2019)\citenamefont {Wang},
  \citenamefont {Hou}, \citenamefont {Lu}, \citenamefont {Chen}, \citenamefont
  {Zhang},\ and\ \citenamefont {Chan}}]{wang2019arbitrary}%
  \BibitemOpen
  \bibfield  {author} {\bibinfo {author} {\bibfnamefont {S.}~\bibnamefont
  {Wang}}, \bibinfo {author} {\bibfnamefont {B.}~\bibnamefont {Hou}}, \bibinfo
  {author} {\bibfnamefont {W.}~\bibnamefont {Lu}}, \bibinfo {author}
  {\bibfnamefont {Y.}~\bibnamefont {Chen}}, \bibinfo {author} {\bibfnamefont
  {Z.}~\bibnamefont {Zhang}},\ and\ \bibinfo {author} {\bibfnamefont {C.~T.}\
  \bibnamefont {Chan}},\ }\href@noop {} {\bibfield  {journal} {\bibinfo
  {journal} {Nature communications}\ }\textbf {\bibinfo {volume} {10}},\
  \bibinfo {pages} {832} (\bibinfo {year} {2019})}\BibitemShut {NoStop}%
\bibitem [{\citenamefont {Wang}\ \emph {et~al.}(2020)\citenamefont {Wang},
  \citenamefont {Wang},\ and\ \citenamefont {Hu}}]{wang2020waveguide}%
  \BibitemOpen
  \bibfield  {author} {\bibinfo {author} {\bibfnamefont {X.-Y.}\ \bibnamefont
  {Wang}}, \bibinfo {author} {\bibfnamefont {F.-F.}\ \bibnamefont {Wang}},\
  and\ \bibinfo {author} {\bibfnamefont {X.-Y.}\ \bibnamefont {Hu}},\
  }\href@noop {} {\bibfield  {journal} {\bibinfo  {journal} {Physical Review
  A}\ }\textbf {\bibinfo {volume} {101}},\ \bibinfo {pages} {053820} (\bibinfo
  {year} {2020})}\BibitemShut {NoStop}%
\bibitem [{\citenamefont {Wiersig}(2022)}]{wiersig2022revisiting}%
  \BibitemOpen
  \bibfield  {author} {\bibinfo {author} {\bibfnamefont {J.}~\bibnamefont
  {Wiersig}},\ }\href@noop {} {\bibfield  {journal} {\bibinfo  {journal}
  {Physical Review A}\ }\textbf {\bibinfo {volume} {106}},\ \bibinfo {pages}
  {063526} (\bibinfo {year} {2022})}\BibitemShut {NoStop}%
\bibitem [{\citenamefont {Azzam}\ \emph
  {et~al.}(2018{\natexlab{b}})\citenamefont {Azzam}, \citenamefont {Shalaev},
  \citenamefont {Boltasseva},\ and\ \citenamefont
  {Kildishev}}]{azzam2018formation}%
  \BibitemOpen
  \bibfield  {author} {\bibinfo {author} {\bibfnamefont {S.~I.}\ \bibnamefont
  {Azzam}}, \bibinfo {author} {\bibfnamefont {V.~M.}\ \bibnamefont {Shalaev}},
  \bibinfo {author} {\bibfnamefont {A.}~\bibnamefont {Boltasseva}},\ and\
  \bibinfo {author} {\bibfnamefont {A.~V.}\ \bibnamefont {Kildishev}},\
  }\href@noop {} {\bibfield  {journal} {\bibinfo  {journal} {Physical review
  letters}\ }\textbf {\bibinfo {volume} {121}},\ \bibinfo {pages} {253901}
  (\bibinfo {year} {2018}{\natexlab{b}})}\BibitemShut {NoStop}%
\bibitem [{\citenamefont {Yang}\ \emph {et~al.}(2017)\citenamefont {Yang},
  \citenamefont {Miller}, \citenamefont {Christensen}, \citenamefont
  {Joannopoulos},\ and\ \citenamefont {Soljacic}}]{yang2017low}%
  \BibitemOpen
  \bibfield  {author} {\bibinfo {author} {\bibfnamefont {Y.}~\bibnamefont
  {Yang}}, \bibinfo {author} {\bibfnamefont {O.~D.}\ \bibnamefont {Miller}},
  \bibinfo {author} {\bibfnamefont {T.}~\bibnamefont {Christensen}}, \bibinfo
  {author} {\bibfnamefont {J.~D.}\ \bibnamefont {Joannopoulos}},\ and\ \bibinfo
  {author} {\bibfnamefont {M.}~\bibnamefont {Soljacic}},\ }\href@noop {}
  {\bibfield  {journal} {\bibinfo  {journal} {Nano letters}\ }\textbf {\bibinfo
  {volume} {17}},\ \bibinfo {pages} {3238} (\bibinfo {year}
  {2017})}\BibitemShut {NoStop}%
\end{thebibliography}%

\end{document}